\documentclass[sigconf,screen,nonacm]{acmart} 
\pdfoutput=1
\usepackage[english]{babel}
\usepackage[utf8]{inputenc}
\usepackage[capitalize]{cleveref}
\usepackage[shortlabels]{enumitem}
\usepackage{tikz-cd}
\usepackage{titlesec}
\usepackage{fixes}
\usepackage{theorems}
\usepackage{macros}

\newif\ifappendix
\appendixtrue
\newcommand{\onappendix}[1]{\ifappendix{#1}\fi}

\usetikzlibrary{calc}

\setcopyright{acmlicensed}
\copyrightyear{2018}
\acmYear{2018}
\acmDOI{XXXXXXX.XXXXXXX}

\acmConference[LICS '24]{39th Annual ACM/IEEE Symposium on Logic in Computer Science (LICS)}{June 03--05, 2018}{Woodstock, NY}
\acmISBN{978-1-4503-XXXX-X/18/06}

\newcommand{\PCGpd}{\operatorname{PCGpd}}

\begin{document}
\title{Delooping cyclic groups with lens spaces in homotopy type theory}
\author{Samuel Mimram}
\orcid{0000-0002-0767-2569}
\email{samuel.mimram@polytechnique.edu}
\author{Émile Oleon}
\orcid{0009-0001-8398-2577}
\email{emile.oleon@polytechnique.edu}
\affiliation{
  \institution{LIX, CNRS, École polytechnique, Institut Polytechnique de Paris}
  \streetaddress{91120}
  \city{Palaiseau}
  \country{France}
}

\begin{abstract}
  In the setting of homotopy type theory, each type can be interpreted as a
space. Moreover, given an element of a type, i.e. a point in the corresponding
space, one can define another type which encodes the space of loops based at
this point. In particular, when the type we started with is a groupoid, this
loop space is always a group. Conversely, to every group we can associate a type
(more precisely, a pointed connected groupoid) whose loop space is this group:
this operation is called \emph{delooping}. The generic procedures for
constructing such deloopings of groups (based on torsors, or on descriptions of
Eilenberg-MacLane spaces as higher inductive types) are unfortunately equipped
with elimination principles which do not directly allow eliminating to
untruncated types, and are thus difficult to work with in practice. Here, we
construct deloopings of the cyclic groups~$\Z_m$ which are \emph{cellular}, and
thus do not suffer from this shortcoming. In order to do so, we provide
type-theoretic implementations of \emph{lens spaces}, which constitute an
important family of spaces in algebraic topology. Our definition is based on the
computation of an iterative join of suitable maps from the circle to an
arbitrary delooping of $\Z_m$. In some sense, this work generalizes the
construction of \emph{real projective spaces} by Buchholtz and Rijke, which
handles the case $m=2$, although the general setting requires more involved
tools. Finally, we use this construction to also provide cellular descriptions
of dihedral groups, and explain how we can hope to use those to compute the
cohomology and higher actions of such groups.


\end{abstract}

\maketitle


\section{Introduction}
Homotopy type theory (HoTT), which is based on
Martin-Löf type theory~\cite{martin1984intuitionistic},
was introduced around 2010~\cite{hottbook}. It stems from the idea
that types in logic should be interpreted not as sets, as traditionally
done in the semantics of logic, but rather as \emph{homotopy types}, by which we
mean spaces considered up to deformation. The identities between two
elements of a type can then be thought of as paths between points corresponding to
the elements~\cite{awodey2009homotopy}, and the fact that two identities between
the same points are not necessarily the same~\cite{hofmann1998groupoid} means
that types can bear a non-trivial geometry. This point of view, which is
validated by Voevodsky's simplicial model of univalent type
theory~\cite{kapulkin2021simplicial}, allows one to prove properties about
spaces by reasoning within type theory. This is satisfactory from a practical
point of view, because we can use proof assistants to check our proofs in full details,
but also from a theoretical point of view because constructions performed in
type theory are homotopy invariant by construction. This last point is actually
both a blessing and a curse: it often means that the traditional proofs have to
be deeply reworked in order to have a chance to be mechanized.


In this article, we provide a construction of \emph{lens spaces}, which
constitute an important family of spaces. Moreover, our construction is
performed by iterating pushout constructions. They can thus be thought of as a
type-theoretic counterpart of their description as cell complexes, and therefore
come equipped with a recursion principle which allows eliminating to untruncated
types.

\subsection{Delooping groups}
The motivation for defining those spaces stems from the desire to perform
computations on groups in homotopy type theory. In this setting, every pointed
type~$A$ has a loop space~$\Loop A$, which is a group when $A$ is a
groupoid. Conversely, any group~$G$ arises in this way: we can always define a
pointed space~$\B G$ whose loop space is~$G$, which is the reason why $\B G$ is
called a \emph{delooping} of~$G$. Moreover, this construction induces an
equivalence between groups and pointed connected spaces. This construction can
be seen as a way to encode a group structure into a space, while retaining its
main properties; for instance, the cohomology of $\B G$ as a space is the
traditional cohomology of~$G$ as a group.

There are very general constructions for delooping groups. For instance, those
are a particular case of Eilenberg-MacLane spaces, which can be explicitly
constructed as higher inductive types in HoTT~\cite{licata2014eilenberg}. There
is however an important limitation of such constructions, which make them very
difficult to use for some computations: they produce \emph{recursive} higher
inductive types (or HITs), due to the use of truncations. More explicitly, given
a group~$G$, its delooping can be constructed as the HIT generated by
\begin{enumerate}
\item[(0)] one point~$\star$,
\item[(1)] one identity $[a]:\star=\star$ for every element $a$
of the group~$G$,
\item[(2)] one relation $[a]\cdot[b]=[ab]$ for every pair of elements $a$
  and~$b$ of the group, as well as a relation $[1]=1$,
\item[(3)] taking the groupoid truncation of the resulting type.
\end{enumerate}
Because of the last step, we can only easily eliminate to groupoids and we
cannot perform computations by induction of the dimension of the constructors,
which would be possible if we had a description of the delooping as a
non-recursive HIT, what we could call here a \emph{cellular} definition.
Such a cellular structure would provide an elimination principle for~$\B G$
allowing us to effectively construct maps $\B G\to X$ for general types~$X$, not
necessarily groupoids.

\subsection{Constructing cellular deloopings}
In order to produce a cellular definition of the delooping of a given
group~$G$, one can begin from the above description without performing the last
truncation step (3). The pointed connected type $\B_1G$ thus defined is not in
general a groupoid.
In order to obtain one, the general idea consists in progressively adding higher
identity generators to the type, in order to make it more and more connected and
construct a family of types $\B_nG$ without higher-dimensional ``holes'' in
dimensions $1<k\leq n$. Taking the sequential colimit, we will obtain a space
$\B_\infty G$ which is a groupoid and thus a delooping of~$G$.
%
%
The reader familiar with techniques in algebraic topology will observe that this
process is very similar to the one of computing resolutions of algebraic
structures or spaces.

The way we can perform this construction in general is however not clear at
first: we need to find enough cells to provably fill in all the holes in each
dimension, and we need to be able to perform this in a systematic way. An
approach based on rewriting and a generalization of Squier's theorem was
proposed in~\cite{kraus2022rewriting}, but it is quite limited: as for now, it
is only able to perform the very first steps of the above construction, it is
quite involved computationally, and it generally produces types with are not
minimal in terms of number of generators.
Another way to proceed consists in taking inspiration from geometric
constructions where some models of deloopings are known. For instance, the
infinite-dimensional real projective space $\RP\infty$ is known to be a delooping of $\Z_2$
and, moreover, we know a cellular decomposition of this space with one cell in
every dimension. We thus can hope to have a description of it in homotopy
type theory as a cellular complex with one generating cell in every dimension. In fact, this task
was successfully performed by Buchholtz and Rijke in~\cite{buchholtz2017real}.

\subsection{Defining lens spaces}
Here, we further explore the latest route and compute a model for $\B\Z_m$ by
using a familly of spaces well known in traditional algebraic topology, the
\emph{lens spaces}, which are due to Tietze~\cite{tietze1908topologischen}.
Those have been thoroughly studied in low dimension~\cite{watkins} (\eg their homotopy and homeomorphism types are fully classified),
and happen to provide (counter)examples to various conjectures and theorems (for instance, some particular lens spaces do not have the same homotopy type but share the same homotopy and homology groups).
Our main contribution is to define types corresponding to those spaces, and thus
derive a cellular model of~$\B\Z_m$ with infinite-dimensional lens spaces.

Contrarily to what it might first seem, the task is more difficult than simply
replacing $2$ by $m$ in the paper defining projective spaces.
The aforementioned article presents very well the situation with
respect to~$\Z_2$, but it was unclear (at least for us) that the techniques
developed there could be reused in other models. We explain that the
construction performed there is a particular instance of a very general
construction based on tools developed in Rijke's PhD
thesis~\cite{rijke2018classifying}, which generalizes one due to
Milnor~\cite{milnor1956construction, milnor1956construction2}. Basically, the
idea in order to construct a cellular model for $\B G$ consists in starting with
a map $f:X\to\B G$ which is surjective up to homotopy, and compute its iterated
join products~\cite{rijke2017join} in order to produce another definition of the
delooping which is cellular. In fact, it is so general that it can be performed
without starting from a particularly concrete model of the delooping~$\B G$ as
target for~$f$. We also take the opportunity of this paper to rework some of the
associated proofs and provide ones based on the flattening lemma (as opposed to
descent), which eases mechanization.
Another important point is that the construction of lens spaces does not actually start from the same
data as projective spaces, and is not even a generalization of the latter: it is
based on a map $S^1\to\B\Z_m$ rather than $1\to\B\Z_m$, which gives more
latitude to the construction (there are actually multiple choices of such maps
one could take, and the resulting computations are more involved than in the
real projective case).

We believe that the resulting definition of $\B\Z_m$ will be useful in order to
perform computations with cyclic groups, such as computing their cohomology~\cite{buchholtz2020cellular}, as we will explain later. It
should also allow defining actions of~$\Z_m$ on higher types (as opposed to
sets). We leave these applications for future work, as well as the
construction of the delooping of other classical groups.


\subsection{Plan of the paper}
After recalling some basic notions and constructions in homotopy type theory
(\cref{hott}), we introduce the notion of delooping of a group
(\cref{deloopings}). We then define and study the join operation on types and
morphisms, which allows computing the image of a map
(\cref{milnor-construction}), and can be reformulated as an operation on fiber
sequences (\cref{fiber-sequences}). These constructions, along with a
correspondence between fiber sequences and group actions, allow us to define a
type-theoretic counterpart of lens spaces, and thus a model of $\B\Z_m$
(\cref{lens-spaces}), which is shown to be cellular (\cref{cellularity}). We
then apply this work in order to define deloopings of dihedral groups
(\cref{delooping-dihedral}) and conclude (\cref{conclusion}).


\section{Homotopy type theory}
\label{hott}
Throughout the article, we suppose the reader is already familiar with homotopy
type theory and refer to the book~\cite{hottbook} for reference. We only fix
here some notations for the classical notions used in the article.

\subsection{Types}
We write~$\U$ for the universe, whose elements are types which are small (for
simplicity, we do not detail universe levels throughout the article). We write
$x:A$ to indicate that $x$ is an element of a type~$A$. The initial and terminal
types are respectively denoted by~$0$ and~$1$.

Given two types~$A$ and~$B$, we write $A\to B$ for the type of functions between
them. A function $f:A\to B$ is an \emph{equivalence} when it admits both a left and a right inverse
up to homotopy. We write~$A\equivto B$ for the type of equivalences between~$A$
and~$B$.
Given a type~$A$ and a type family $B:A\to\U$, we write $(x:A)\to B(x)$ or
$\Pi(x:A).B(x)$ or $\Pi A.B$ for the induced \emph{dependent product} type, and
$\Sigma(x:A).B(x)$ or $\Sigma A.B$ for the induced \emph{dependent sum}
type. The two canonical projections are noted $\fst:\Sigma A.B\to A$ and
$\snd:(x:\Sigma A.B)\to B(\fst(x))$. Given morphisms $f:A\to A'$ and
$g:(x:A)\to B(x)\to B'(f(x))$, we write $\Sigma f.g:\Sigma A.B\to\Sigma A'.B'$
for the canonically induced morphism.

\subsection{Paths}
We write $x\defd y$ to indicate that $x$ is \emph{defined} to be $y$. Our type
theory also features a type corresponding to identifications between elements and, given two elements
$x$ and $y$ of a type $A$, we write $x=y$ for their identity type. In
particular, an element of $x=x$ is called a \emph{loop} on~$x$. Any element~$x$
induces a canonical loop $\refl[x]:x=x$. Given paths $p:x=y$ and $q:y=z$, we write
$p\pcomp q:x=z$ for their concatenation and $\pinv{p}:y=x$ for the inverse
of~$p$.

Any path $p:A=B$ between types $A$ and $B$, induces a function
$\transport p:A\to B$, called \emph{transport} along $p$, which is an
equivalence whose inverse is denoted $\transportinv p:B\to A$. The
\emph{univalence} axiom states that the map $(A=B)\to(A\equivto B)$ thus induced
is an equivalence.

By \emph{congruence}, any function~$f:A\to B$ and path $p:x=y$ induce an
equality $\ap fp:f(x)=f(y)$. In particular, given a type family~$B:A\to\U$ and a
path $p:x=y$ in~$A$, we have an induced equality $B(x)=B(y)$ and thus an
equivalence $B(x)\equivto B(y)$ whose underlying function is noted
$\subst Bp:B(x)\to B(y)$ (and $\substinv Bp$ for its inverse) and corresponds to
\emph{substituting} $x$ by~$y$ in~$B$.

Given a type family $B:A\to\U$, a path $p:x=y$, and elements $x':B(x)$ and
$y':B(y)$, we write $x'=^B_py'$ (or even $x'=_py'$ when $B$ is clear from the
context) for the type of \emph{paths over} $p$ from~$x'$ to~$y'$; by definition,
this type is the identity type $\subst Bp(x')=y'$.

Given functions $f,g:A\to B$, a path $p:f=g$ between those, and $x:A$, we write
$\happly px:f(x)=g(x)$ for the induced path obtained by applying~$p$ to~$x$
pointwise.

\subsection{Homotopy levels}
A type~$A$ is \emph{contractible} when the type $\Sigma(x:A).\Pi(y:A).(x=y)$ is
inhabited. A contractible type is also called a \emph{$(-2)$-type}, and we
define by induction an $(n+1)$-type to be a type~$A$ such that $x=y$ is an
$n$-type for every elements $x,y:A$. In particular, $(-1)$-, $0$- and $1$-types
are respectively called \emph{propositions}, \emph{sets} and \emph{groupoids}.
Given a type~$A$, we write $\trunc nA$ for the \emph{$n$-truncation} of~$A$,
which is the universal way of turning a type into an $n$-type.
A type~$A$ is \emph{connected} when $\trunc 0A$ is contractible ($\trunc 0A$ being the set of connected components of $A$).

\subsection{Spheres}
We write $\S n$ for the type corresponding to the $n$-sphere (those can for
instance be defined inductively by $\S{-1}=0$ and $\S{n+1}=\susp \S n$, where
$\susp$ is the suspension operation). For $\S1$, we write $\pt{}$ for the
canonical element and $\Sloop:\pt{}=\pt{}$ for the canonical non-trivial loop.

\subsection{Homotopy limits and colimits}
\label{homotopy-limits}
All (co)limits considered in the paper are homotopic. Given two maps $f,g$ with
the same codomain, we have a pullback square
\[
  \begin{tikzcd}
    \Sigma(x,y:A\times B).(f(x)=g(y))\ar[d,"\fst\circ\snd"']\ar[r,"\fst"]\ar[dr,phantom,pos=0,"\lrcorner"]&A\ar[d,"f"]\\
    B\ar[r,"g"']&C
  \end{tikzcd}
\]
As a particular case, when $A$ and $B$ are the terminal type and
$f:1\to C$ and $g:1\to C$ are maps respectively pointing at elements $x$ and
$y$, their pullback is the identity type $x=y$.


\section{Deloopings of groups}
\label{deloopings}

\subsection{Deloopings}
A \emph{group}~$G$ is a set equipped with multiplication $m:G\to G\to G$ and
unit $e:1\to G$ satisfying the usual axioms, and a morphism of groups is a
function between the underlying sets preserving the operations.

A \emph{pointed type} $A$ consists of a type $A$ together with an element
$\pt{}:A$, and a pointed map $f:A\pto B$ consists of a function $f:A\to B$
together with a equality $\pt f:f(\pt A)=\pt B$. The \emph{loop space} of a
pointed type~$A$ is $\Loop A:=(\pt{}=\pt{})$, the type of loops on the
distinguished element.
This operation is functorial: any pointed map $f:A\to B$ canonically induces a
map $\Loop f:\Omega A\to\Omega B$~\cite[Definition~8.4.2]{hottbook} defined as
$\Loop f(p)=\pinv{\pt f}\pcomp \ap fp\pcomp\pt f$, and this construction is
compatible with composition and identities. When~$A$ is a groupoid (in the sense
of a $1$-truncated type), $\Loop A$ is thus a group with multiplication given by
concatenation of paths and unit given by the constant path.

A \emph{delooping} of a group~$G$ is a pointed connected type $\B G$ equip\-ped
with an equality of groups $\Beq G:\Loop\B G=G$. Such a type is necessarily a
groupoid since~$G$ is a set.
%
%
The following lemma states that we can also deloop morphisms,
see~\cite[Section~4.10]{symmetry} and~\cite[Corollary~12]{warn2023eilenberg}, as
well as the appendix, for a proof:

\begin{lemma}
  \label{delooping-morphisms}
  Given two pointed connected groupoids~$A$ and~$B$, and a group morphism
  $f:\Loop A\to\Omega B$, there is a unique pointed morphism $g:A\pto B$ such
  that~$\Loop g=f$.
\end{lemma}

\noindent
Given a morphism of groups $f:G\to H$, where $G$ and $H$ admit deloopings, the
\emph{delooping} of~$f$ is the map
\[
  \B f:\B G\to\B H
\]
associated, by \cref{delooping-morphisms}, to the morphism obtained as the
composite
\[
  \begin{tikzcd}
    \Loop\B G\ar[r,"\transport{\Beq G}"]&G\ar[r,"f"]&H\ar[r,"\transportinv{\Beq H}"]&\Omega\B H\text{.}
  \end{tikzcd}
\]
This operation is easily shown to be functorial\onappendix{ (\cref{delooping-functorial})} in that it preserves compositions and identities.

The delooping of a given group~$G$ is unique in the sense that any two
deloopings of a given group are necessarily equal, thus justifying the notation
$\B G$\onappendix{ (see \cref{delooping-unique} for a more precise statement and proof)}. It
can also be shown that any group~$G$ always admits a delooping:

\begin{lemma}
  \label{deloopings-exist}
  Every group~$G$ admits a delooping $\B G$.
\end{lemma}

\noindent
There are (at least) two generic ways of constructing $\B G$ for an
arbitrary~$G$. We only briefly mention those here because we will only need to
know that the above lemma is true, but will not depend on a particular such
construction.
The first one consists in defining $\B G$ as a truncated HIT, as described in
the introduction: this is obtained as the $K(G,1)$ construction, defined by
Finster and Licata in~\cite{licata2014eilenberg}.
The second one is the \emph{torsor} construction: it consists in defining $\B G$
as the connected component in $G$-sets of the principal $G$-set (a $G$-set is a
set equipped with an action of~$G$). More details can be found
in~\cite{symmetry}\onappendix{ or \cref{delooping-with-G-sets}}.
Useful smaller variants of this construction can also be
considered~\cite{delooping-generated}.
Finally, we should mention that interesting models for deloopings can often be
constructed for particular groups. For instance, one of the main early results
of homotopy type theory is that the circle $\S1$ is a delooping
of~$\Z$~\cite[Section~8.1]{hottbook}. New examples are given in the present
paper.

The fundamental theorem about the delooping construction is the following:

\begin{theorem}
  \label{groups-vs-pointed-connected-groupoids}
  The functions $\Loop$ and $\B$ induce an equivalence between the category of
  groups and the category of pointed connected groupoids.
\end{theorem}

\noindent
The above proposition states that a group is the same as a pointed connected
groupoid: the first notion can be thought of as an \emph{external} one (in the
sense that we impose axioms on the structures), whereas the second is an
\emph{internal} one (in the sense that the structure is generated by the
properties of the considered types).
%
%
%
It also provides an elimination principle for deloopings. Namely, functions
$\B G\to A$ correspond to group morphisms $G\to\Loop A$ when $A$ is a pointed
connected groupoid. We could easily drop the restriction to connected types, by
considering connected components. However, we do not have a direct way of
eliminating to types which are not groupoids for now: we will address this point
with our construction.

\subsection{Actions of groups}
\label{group-action}
\label{homotopy-quotient}
Given a group~$G$ and a type $A$, an \emph{action} of~$G$ on $A$ is a map $f:\B G\to\U$ such that $f(\pt{})=A$. Namely, writing $A\defd f(\pt{})$, each path $a:\pt{}=\pt{}$ (which can be seen as an element of~$G$), induces by substitution an equivalence $A\to A$, in a way which is compatible with composition and identities. In fact, it can be shown
that actions on sets in the above sense correspond to actions of groups in the
traditional sense\onappendix{ (\cref{internal-group-action})}.

Given an action $f:\B G\to\U$, with $A\defd f(\star)$, the \emph{homotopy
  quotient} of $A$ by~$G$ (with respect to the action) is the type
\[
  A\hq G
  \defeq
  \Sigma(x:\B G).f(x)
\]
There are canonical \emph{quotient} and \emph{projection} morphisms,
respectively defined as
\begin{align*}
  \qtt:A&\to A\hq G
  &
  \fst:A\hq G&\to\B G
  \\
  x&\mapsto(\pt{},x)
  &
  (x,y)&\mapsto x
\end{align*}
where the quotient~$\qtt$ witnesses the inclusion of~$A$ as the fiber
above~$\pt{}$ in the homotopy quotient.
%
%


\section{The Milnor construction}
\label{milnor-construction}
The \emph{Milnor construction} is a general method, due to
Milnor~\cite{milnor1956construction,milnor1956construction2}, to construct the
universal bundle for a fixed topological group~$G$. Its adaptation to the
setting of homotopy type theory was done by Rijke in his PhD
thesis~\cite{rijke2017join,rijke2018classifying}.

\subsection{The join construction}
Given two types $A$ and $B$, their \emph{join} $A\join B$,
see~\cite[Section~6.8]{hottbook}, is the pushout of the product projections:
\[
  \begin{tikzcd}
    A\times B\ar[d,"\fst"']\ar[r,"\snd"]\ar[dr,phantom,pos=1,"\ulcorner"]&B\ar[d,dotted,"\sndinj"]\\
    A\ar[r,dotted,"\fstinj"']&A\join B
  \end{tikzcd}
\]
As any pushout, the join can be constructed as a quotient of the
coproduct~$A\sqcup B$, and this description provides a way to implement the join
of two types as a higher inductive type: $A\join B$ is the universal type which
contains every element of~$A$, every element of~$B$, and has a path between any
element of~$A$ and any element of~$B$. For instance, the space corresponding
to~$\S0\join\S0$ looks like
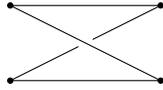
\[
  \begin{tikzpicture}
    \filldraw (0,1) circle (.035);
    \filldraw (0,0) circle (.035);
    \filldraw (2,1) circle (.035);
    \filldraw (2,0) circle (.035);
    \draw (0,0) -- (2,1);
    \draw (0,0) -- (2,0);
    \draw (0,1) -- (2,1);
    \filldraw[white] (1,.5) circle (.1);
    \draw (0,1) -- (2,0);
  \end{tikzpicture}
\]
and is equal to $\S1$. Indeed, for a type $A$, we have that $\S0\join A$
is the suspension $\susp A$ so that $\S0\join \S0=\susp\S0=\S1$ and more
generally $\S m\join \S n=\S{m+n+1}$. The join operation is associative,
commutative, and admits the empty type~$0$ as neutral
element~\cite[Section~1.8]{brunerie2016homotopy}.

\subsection{Propositional truncation}
One of the interests of the join construction is that it results in a type which is
more connected than its operands. Namely, recall that a type $A$ is
\emph{$n$-connected} when its $n$-truncation $\trunc nA$ is contractible. It can
be shown that when $A$ is $m$-connected and $B$ is $n$-connected their join
is $(m+n+2)$-connected~\cite[Proposition~3]{brunerie2019james} (in the above
example, $\S1$ is $0$-connected, as the join of two $(-1)$-connected types). If
we iterate this construction by computing the join product $A^{\join n}$ of $n$
copies of~$A$ for $n$ increasing, we obtain more and more connected spaces: if $A$
is $k$-connected then $A^{\join n}$ is $(n(k{+}2){-}2)$-connected. The map
$\fstinj$ of the pushout provides us with a canonical inclusion of $A^{\join n}$
into $A^{\join(n+1)}$ and we write $A^{\join\infty}$ for the colimit of the
diagram
\[
  \begin{tikzcd}
    0\ar[r]&A\ar[r]&A\join A\ar[r]&A\join A\join A\ar[r]&\cdots
  \end{tikzcd}
\]
We expect this limit to be $\infty$-connected, and even contractible, excepting in
one case: if we begin with the empty type, the limit will still be the empty
type. The proper way to phrase this is as
follows~\cite[Theorem~4.2.7]{rijke2018classifying}:

\begin{theorem}
  The type $A^{\join\infty}$ is the propositional truncation of~$A$.
\end{theorem}



\noindent
In particular, with $A\defd\S0$, we deduce that the infinite-dimensional
sphere $\S\infty:=(\S0)^{\join\infty}$ is contractible.

\subsection{Join of maps}
The join construction can be generalized to morphisms as follows. Given
maps~$f:A\to C$ and $g:B\to C$ with the same target, their \emph{join} $f\join g$ is the universal map
from the pushout of the dependent projections~$\fst$ and~$\snd$:
\[
  \begin{tikzcd}
    A\times_CB\ar[d,"\fst"']\ar[r,"\snd"]&B\ar[d,"\sndinj"]\ar[ddr,bend left,"g"]\\
    A\ar[drr,"f"',bend right=20]\ar[r,"\fstinj"']&A\join_CB\ar[dr,"f\join g"description]\\
    &&C
  \end{tikzcd}
\]
The source is abusively noted $A\join_CB$ and we have $A\join_1B=A\join B$, so
that this generalizes the previous construction which consisted of taking the join of maps to the terminal object.

\subsection{Fibers of maps}
\label{fiber}
Given a function $f:A\to B$ and a point $b:B$, the \emph{fiber} of~$f$ at~$b$ is
\[
  \fib f(b)=\Sigma(x:A).(f(x)=b)  
\]
This type can also be seen as the pullback of the constant arrow at~$b$
along~$f$:
\[
  \begin{tikzcd}
    \fib f(b)\ar[d,dotted,"i"']\ar[r,dotted]\ar[dr,phantom,pos=0,"\lrcorner"]&1\ar[d,"b"]\\
    A\ar[r,"f"']&B
  \end{tikzcd}
\]
where the map $i$ is called here the canonical inclusion.
In particular, given a pointed type~$B$ and a function $f:A\to B$, we define the
\emph{kernel} of $f$ as $\ker f=\fib f(\pt B)$,
see~\cite[Section 8.4]{hottbook}. When $A$ is pointed and $f$ is a pointed map,
$\ker f$ is canonically pointed (by $\pt{A}$ together with the proof of
$f(\pt A)=\pt B$ given by the fact that $f$ is pointed).

It can be shown that the join operation commutes with taking fibers, in the
following sense~\cite[Theorem~2.3.15]{rijke2018classifying}:

\begin{theorem}
  \label{thm:fib-join}
  \label{fib-join}
  Given $f:A\to X$ and $g:B\to X$ and $x:X$, we have
  \[
    \fib{f\join g}(x)=\fib f(x)\join\fib g(x)
    \text.
  \]
\end{theorem}

\noindent
We provide a proof of a generalization of this result in \cref{cellularity}.

\subsection{Image of maps}
If we consider the iterated join $f^{\join n}$ of $n$ instances of a
map~$f:A\to B$, its fibers get more and more connected as $n$ increases, since
$\fib{f^{\join n}}(b)=\fib{f}(b)^{\join n}$. If we take the colimit
$f^{\join\infty}$, we obtain a map whose fibers are propositions: this is the
canonical inclusion of the image of~$f$ into the codomain of~$f$. More
precisely, writing $\im^n(f)$ for the source of $f^{\join n}$, we have a
canonical map $\im^{n}(f)\to\im^{n+1}(f)$ given by $\fstinj$, and we write
$\im^\infty(f)$ for the colimit of the diagram
\[
  \begin{tikzcd}
    \im^0(f)\ar[r]&\im^1(f)\ar[r]&\im^2(f)\ar[r]&\cdots
  \end{tikzcd}
\]
The maps $f^{\join n}:\im^n\to B$ form a cocone on this diagram and, by
universal property, we obtain a map $i:\im^\infty(f)\to B$. It can then be
shown, see \cite[Theorem~4.2.13]{rijke2018classifying} and
\cite[Theorem~3.3]{rijke2017join}, that $\im^\infty(f)$ coincides with the usual
definition of the image of a morphism $\im(f)=\Sigma(y:B).\ptrunc{\fib f(y)}$,
see~\cite[Definition 7.6.3]{hottbook}:

\begin{theorem}
  \label{colimit-image}
  We have $\im^\infty(f)=\im(f)$ and $i$ is the canonical projection of the
  image on the first component.
\end{theorem}

In the case we start from a map $f:A\to B$ which is surjective, in the sense
that we have $\ptrunc{\fib f(y)}$ for every $y:B$~\cite[Definition
4.6.1]{hottbook}, we have $\im(f)=B$, and the colimit of~$\im^n(f)$ is~$B$. Each
$\im^n(f)$ is obtained from the previous one by a pushout construction and we
thus recover well-known cell complexes approximation for spaces in homotopy type theory (see \cref{ex:RP} below).
In practice, the following result is often useful to show that a map is surjective:

\begin{lemma}
  \label{pointed-connected-surjective}
  Given types $A$ and $B$, with $A$ merely inhabited and $B$ connected, any map $f:A\to B$ is surjective.
\end{lemma}
\begin{proof}
  Since being surjective is a proposition, we can suppose given a distinguished element~$a$ of~$A$.
  Given $b:B$, we know that there merely exists a path $p:f\,a=b$, which we can use to show the
  proposition $\ptrunc{\fib f(b)}$. Namely, the type $\ptrunc{\fib f(f(a))}$ is
  inhabited (by~$a$) and we conclude by transport along~$p$.
\end{proof}

\begin{example}[from~\cite{buchholtz2017real}]
  \label{ex:RP}
  \label{RP}
  Consider the canonical map $f:1\to\B\Z_2$. Since $\B\Z_2$ is connected, this
  map is surjective by \cref{pointed-connected-surjective}.
  Given $n\in\N$, we define the $n$-th \emph{real projective space} as
  $\RP n=\im^{n+1}(f)$ and the associated \emph{tautological line bundle}
  $\ell^n:\RP n\to\B\Z_2$ as $\ell^n=f^{\join(n+1)}$ (the shift of indices was
  chosen in order to match usual conventions). Explicitly, this means that we
  take $\RP{-1}=\im^0(f)=0$ and define $\RP{n+1}$ as the pushout
  \[
    \begin{tikzcd}
      \RP n\times_{B\Z_2}1\ar[d]\ar[r]\ar[dr,phantom,"\ulcorner",pos=1]&1\ar[d]\ar[ddr,bend left,"f"]\\
      \RP n\ar[drr,bend right=20,"\ell^n"']\ar[r]&\RP{n+1}\ar[dr,"\ell^{n+1}=\ell^n\join f"description]\\
      &&B\Z_2
    \end{tikzcd}
  \]
  where, by definition of the kernel we have
  $
  \RP n\times_{B\Z_2}1=\ker\ell^n
  $.
  Since $\ell^{n+1}=\ell^n\join f$, by \cref{thm:fib-join}, we have
  $\ker\ell^{n+1}=\ker\ell^n\join\ker f$, and we thus have a pullback
  \[
    \begin{tikzcd}
      \ker{\ell^n}\join\ker{f}\ar[d]\ar[r]\ar[dr,phantom,"\lrcorner",pos=0]&1\ar[d,"f"]\\
      \RP{n+1}\ar[r,"\ell^{n+1}"']&B\Z_2
    \end{tikzcd}
  \]
  Moreover, $\ker f$ is the pullback of $f$ along itself, which is $\Loop\B\Z_2$
  (see \cref{homotopy-limits}), and thus $\Z_2$ by definition of $\B\Z_2$.
  We have thus shown
  \[
    \ker\ell^{n+1}=\ker\ell^n\join\Z_2
  \]
  Since $\ker\ell^{-1}=0$, and $\Z_2 = \S 0$, it follows by induction that
  \[
    \RP n\times_{B\Z_2}1
    =
    \Z_2^{\join(n+1)}
    =
    \S n
  \]
  To sum up, we have a pushout
  \[
    \begin{tikzcd}
      \S n\ar[d]\ar[r]\ar[dr,phantom,"\ulcorner",pos=1]&1\ar[d]\\
      \RP n\ar[r]&\RP{n+1}
    \end{tikzcd}
  \]  
  If we write $\RP\infty=\colim_n\RP n$ and $\ell^\infty=\colim_n\ell^n$ for the
  colimiting constructions (as described at the beginning of this section), we have
  $\RP\infty=\im f=\B\Z_2$ by \cref{colimit-image}.
\end{example}


\section{Fiber sequences}
\label{fiber-sequences}

\subsection{Definition}
\label{sec:fiber}
A \emph{fiber sequence}
\begin{equation}
  \label{eq:fiber-sequence}
  \begin{tikzcd}
    F\ar[r,hook,"i"]&E\ar[r,->>,"f"]&B
  \end{tikzcd}
\end{equation}
consists of three types $F$ (the \emph{fiber}), $E$ (the \emph{total space}) and
$B$ (the \emph{base space}), with $B$ pointed, and two maps ($i$ and~$f$) such
that $F$ is the kernel of~$f$, with $i$ as canonical inclusion,
\ie such that $i$ is the pullback of the pointing map $b:1\to B$ along~$f$.


\subsection{Homotopy from fiber sequences}
Any fiber sequence induces a long exact sequence of homotopy
groups~\cite[Section~8.4]{hottbook}, which in practice allow for the computation of some homotopy groups when a fiber sequence is known. In this article, we will not need to use these constructions
in full generality; in fact, we will only need to use the following lemma.

\begin{lemma}
  \label{lem:exact-equiv}
  Given a fiber sequence of the form \cref{eq:fiber-sequence}
  with $F$ contractible and $B$ connected, $f$ is an equivalence.
\end{lemma}
\begin{proof}
  Showing that~$f$ is an equivalence is equivalent to showing that~$\fib f(b)$ is
  contractible for any $b:B$~\cite[Section~4.4]{hottbook}. From the fiber
  sequence, we have $\fib f(\pt B)=\ker f=F$ and thus $\fib f(\pt B)$ is
  contractible. Since $B$ is connected, the type $\ptrunc{\pt B=x}$ is
  inhabited, and, since being contractible is a proposition, we can suppose that
  we have a path $p:\pt B=x$. By transport, we deduce that $\fib f(x)$ is
  contractible.
\end{proof}

\noindent
We also have the following variant of the above lemma, with a similar proof:

\begin{lemma}
  \label{lem:exact-connected}
  Given a fiber sequence \cref{eq:fiber-sequence} with $F$ $n$-connected, for
  some $n\in\N$, and $B$ connected, $f$ is $n$-connected.
\end{lemma}

\noindent
We recall that when $f:E\to B$ is $n$-connected (\ie its fibers are
$n$-connected), as in the conclusion of the above lemma, it induces
isomorphisms of homotopy groups $\pi_k(E)\simeq\pi_k(B)$ for every natural
number $k\leq n$~\cite[Corollary~8.4.8]{hottbook}.

\subsection{Delooping short exact sequences}
Given $f:G\to H$ a morphism of groups, we define its \emph{kernel} and
\emph{image} respectively as
\begin{align*}
  \ker f&=\Sigma(x:G).(f(x)=0)
  \\
  \im f&=\Sigma(y:H).\ptrunc{\Sigma(x:G).(f(x)=y)}
\end{align*}
Those are canonically equipped with a group structure induced by the one on~$G$.
We write $i:\ker f\to G$ and $j:\im f\to H$ for the first projections.

\begin{lemma}
  \label{B-ker}
  With the above notations, we have $\ker\B f=\B\ker f$ with $\B i$ as canonical
  inclusion into $\B G$.
\end{lemma}
\begin{proof}
  We have $\ker\B f=\Sigma(x:\B G).(\B f(x)=\pt{\B H})$, which is pointed by
  $(\pt{\B G},p)$ with $p:\B f(\pt{\B G})=\pt{\B H}$ given by the fact that
  $\B f$ is pointed. Thus $\Loop\ker\B f$ is the space of loops on
  $(\pt{\B G},p)$ in this type, \ie paths $q:\pt{\B G}=\pt{\B G}$ equipped with
  a proof of $p=^{\B f(-)=\pt{\B H}}_qp$. This last type is equivalent to
  $\sym{(\ap{(\B f)}q)}\pcomp p=p$ (by \cite[Theorem~2.11.3]{hottbook}), and thus
  to $\ap{(\B f)}q=\refl$ (since paths are invertible). Therefore,
  \[
    \begin{array}{r@{\ }l@{\ }l}
      \Loop\ker\B f&=\Sigma(q:\Omega\B G).(\ap{(\B f)}q=\refl)\\
      &=\Sigma(x:G).(f(x)=0)&=\ker f
    \end{array}
  \]
  from which we deduce $\ker\B f=\B\ker f$ by uniqueness of deloopings. The fact
  that $\B i$ is the canonical inclusion is routine check.
  %
\end{proof}

A diagram of groups
\[
  \begin{tikzcd}
    1\ar[r]&F\ar[r,"f"]&G\ar[r,"g"]&H\ar[r]&1
  \end{tikzcd}
\]
is a \emph{short exact sequence} when $f$ is injective, $\ker g=\im f$, and $g$
is surjective. By delooping, such data induces a fiber sequence, which can be
thought of as stating that the delooping functor is exact:

\begin{lemma}
  \label{delooping-short-exact-sequences}
  Given a short exact sequence as above, the diagram
  \[
    \begin{tikzcd}
      \B F\ar[r,hook,"\B f"]&\B G\ar[r,->>,"\B g"]&\B H
    \end{tikzcd}
  \]
  is a fiber sequence.
\end{lemma}
\begin{proof}
  We have
  \begin{align*}
    \ker(\B g)
    &=\B(\ker g)&&\text{by \cref{B-ker}}\\
    &=\B(\im f)&&\text{by exactness in the middle}\\
    &=\B F&&\text{because $f$ is injective.}\qedhere
  \end{align*}
\end{proof}


\subsection{Join of fiber sequences}
\label{join-of-fiber-sequences}
The join operation also operates on fiber sequences as follows, which allows
reformulating the Milnor construction in a concise way. Any two fiber sequences
$F\into A\overset{f}\onto B$ and $F'\into A'\overset{f'}\onto B'$
induce a new fiber sequence
$F\join F'\into A\join_BA'\onto B$, by the join operation and \cref{thm:fib-join}.
In particular, if we start with a fiber sequence
\[
  \begin{tikzcd}
    F\ar[r,hook]&A\ar[r,->>,"f"]&B
  \end{tikzcd}
\]
and iterate this operation, we obtain a family of fiber sequences
\[
  \begin{tikzcd}
    F^{\join n}\ar[r,hook]&\im^n(f)\ar[r,->>,"f^{\join n}"]&B
  \end{tikzcd}
\]
which converges to the fiber sequence
\[
  \begin{tikzcd}
    \ptrunc{F}\ar[r,hook]&\im(f)\ar[r,->>,"f^{\join\infty}"]&B
  \end{tikzcd}
\]

\begin{proposition}
  \label{infinite-join}
  When $F$ has a point and $B$ is connected, $f^{\join\infty}$ is an equivalence.
\end{proposition}
\begin{proof}
  Since $F$ has a point, we have $\ptrunc{F}=1$, and $f^{\join\infty}$ is thus an
  equivalence by \cref{lem:exact-equiv}.
\end{proof}

\begin{example}
  Consider again the map $f:1\to\B\Z_2$ of \cref{ex:RP}. We have seen that $\ker f=\Z_2$.
  By the above reasoning, we thus have exact sequences
  \[
    \begin{tikzcd}
      \Z_2^{\join(n+1)}\ar[r,hook]&\im^{n+1}(f)\ar[r,->>,"f^{\join(n+1)}"]&\B\Z_2
    \end{tikzcd}
  \]
  which can be rewritten as
  \[
    \begin{tikzcd}
      \S n\ar[r,hook]&\RP n\ar[r,->>,"\ell^n"]&\B\Z_2
    \end{tikzcd}
  \]
  Since $\S n$ is $(n{-}1)$-connected, by \cref{eq:fiber-sequence} we deduce that
  $\ell^n$ is $(n{-}1)$-connected.
  Moreover, by \cref{infinite-join}, we obtain an equality $\RP\infty=\B\Z_2$ at
  the limit.
\end{example}

\subsection{Fibrations and type families}
\label{fibrations-families}
There is a well-known correspondence, due to Grothendieck, between fibrations
and pseudo-presheaves (pseudo-functors from a small category to~$\Cat$), which
admits many variants and extensions (such as the correspondence between discrete
fibrations and presheaves)~\cite{streicher2018fibered}. This type of
correspondence also has an impersonation in the context of type theory as
follows.

Given a type~$B$, a \emph{type over}~$B$ is a type~$A$ together with a map
$f:A\to B$. A \emph{type family} indexed by~$B$ is a function $B\to\U$. We can
define functions between the corresponding types, in both directions:
\[
  \phi:\Sigma(A:\U).(A\to B)\rightleftarrows(B\to\U):\psi
\]
respectively defined by $\phi(A,f)\defd\fib f$, \ie the function $\phi$
associates to any type over~$B$ its fiber, and $\psi(F)\defd(\Sigma B.F,\fst)$,
\ie the function $\psi$ associates to any type family the first projection from
the associated total space. Moreover, these two functions can be checked to form
an equivalence~\cite[Section~4.8]{hottbook}: given $(A,f)$ a type over~$B$, we
have
\[
  \psi(\phi(A,f))=(\Sigma B.\fib f,\fst)=(A,f)
\]
and given a type family $F$ index by~$B$, we have
\[
  \phi(\psi(F))=\fib{\fst}=F
\]
with $\fst:\Sigma B.F\to B$ the projection.

\begin{theorem}
  \label{fibrations-equiv-families}
  For any type~$B$, the types $\Sigma(A:\U).(A\to B)$ and $B\to\U$ are
  equivalent (and thus equal by univalence).
\end{theorem}


Suppose given a group~$G$ and fix $B\defd\B G$. The above theorem then states an
equivalence between morphisms $f:A\to\B G$ and actions of $G$. In particular, a
map $f:E\to\B G$ is the same thing as an action of~$G$ on $\ker f$, and its
homotopy quotient $(\ker f)\hq G$ is~$E$.
%
%
This can also be reformulated as follows in terms of fiber sequences:

\begin{proposition}[action-fibration duality]
  \label{action-fiber-sequence}
  The data of an action of $G$ on a type $X$ whose homotopy quotient is $Y$ is equivalent to the data of a fiber sequence
  \[
    \begin{tikzcd}
      X\ar[r,hook]&Y\ar[r,->>]&\B G
    \end{tikzcd}
  \]
  More precisely, given an action $f:\B G\to\U$ (on $f(\pt{})$) we have a fiber sequence
  \[
    \begin{tikzcd}
      f(\pt{})\ar[r,hook,"\qtt"]&f(\pt{})\hq G\ar[r,->>,"\fst"]&\B G
    \end{tikzcd}
  \]
  where $\qtt$ is the quotient map and $\fst$ is the canonical projection (see
  \cref{homotopy-quotient}). Conversely, given a fiber sequence
  \[
    \begin{tikzcd}
      X\ar[r,hook]&Y\ar[r,->>,"g"]&\B G
    \end{tikzcd}
  \]
  we have the action $\fib g:\B G\to\U$ of $G$ on $X$ (and $Y$ is the homotopy quotient). The two constructions are mutually
  inverse of each other.
\end{proposition}
\begin{proof}
  %
  \newcommand{\FS}{\operatorname{FS}}
  Suppose given two types~$F$ and~$B$. We write $\FS_B$ (\resp $\FS_B^F$) for
  the type of fiber sequences with $B$ as base space (\resp also with $F$ as
  fiber).
  By using the universal property of pullbacks, the type $\FS_B$ can be shown to
  be equivalent to the type $\Sigma(E:\U).(E\to B)$ of types over~$B$, and thus
  to the type of families $B\to\U$ by \cref{fibrations-equiv-families}, see also
  \cite[Proposition~2.3.9]{rijke2018classifying}.
  %
  This proof can then be refined to show that the type $\FS^F_B$ is equivalent
  to the type $\Sigma(f:B\to\U).(f(\pt{})=F)$.
  The desired result then follows by taking $F\defd X$ and $B\defd\B G$.
\end{proof}



\noindent
The join of fiber sequences introduced in \cref{join-of-fiber-sequences}, thus
induces, when the base is $\B G$, an operation which allows combining two
actions $f,g:\B G\to\U$ as an action on $f(\pt{})\join g(\pt{})$
by making the two actions operate coordinatewise.

This Grothendieck correspondence between the two points of view is often useful
(\eg the descent property on fibrations corresponds to the flattening lemma for
type families), and some operations are more easily performed on one side or the
other.
In particular, in our context, instead of using \cref{fib-join}, it will turn
out to be more convenient to consider a dual version, that we prove in
\cref{sigma-join}.

\subsection{Recovering covering spaces}
A classical theorem in algebraic topology states that given an action of a
group~$G$ on a topological space~$X$, which is free and properly discontinuous,
the quotient map $X\to X/G$ is covering with $G$ as
fiber~\cite[Proposition~1.40]{hatcher2009algebraic}. In homotopy type theory,
this translates as the existence of the following fiber sequence, which is a
variant of \cref{action-fiber-sequence} (the freeness and proper discontinuity
conditions do not make sense and are not required, because we are working up to homotopy).

\begin{proposition}
  Given a group~$G$ and an action $f:\B G\to\U$ on a pointed
  type~$X\defd f(\pt{})$, the fiber of the quotient map $q:X\to X\hq G$ is
  $G$, so we have a fiber sequence
  \[
    \begin{tikzcd}
      G\ar[r,hook,"i"]&X\ar[r,->>,"q"]&X\hq G
    \end{tikzcd}
  \]
  where $i$ is the map defined for $x:G$ as
  \[
    i(x)
    \defd
    \subst f{\transportinv{\Beq G}(x)}(\pt{X})
  \]
  which sends an element $x:G$, which can be seen as a path $p:\pt{}=\pt{}$
  of~$\B G$, to the element of~$X$ obtained by transporting the distinguished
  element $\pt{}$ of~$X$ along this path.
\end{proposition}
\begin{proof}
  The homotopy quotient $X\hq G$ is, by definition, $\Sigma(\B G).f$ and the
  quotient map sends $x:X$ to $q(x)\defd(\pt{\B G},x)$. We have
  \begin{align*}
    \fib q(\pt{\B G},\pt{X})
    &=\Sigma(x:X).((\pt{\B G},\pt{X})=q(x))
    \\
    &=\Sigma(x:X).((\pt{\B G},\pt{X})=(\pt{\B G},x))
    \\
    &=\Sigma(x:X).\Sigma(p:\pt{\B G}=\pt{\B G}).(\pt{X}=^f_px)
    \\
    &=\Sigma(x:X).\Sigma(a:G).(i(a)=x)
    \\
    &=\Sigma(a:G).\Sigma(x:X).(i(a)=x)
    \\
    &=\Sigma(a:G).1
    \qquad\qquad\qquad\qquad=G
  \end{align*}
  and the projection map is the projection map from $\Sigma X.\fib i$ to~$X$,
  which is $i$ by \cref{fibrations-equiv-families}.
\end{proof}


\section{Construction of lens spaces}
\label{lens-spaces}

\subsection{Traditional definition of lens spaces}
\label{lens-topology}
In this section, we recall the definition of lens spaces (in the setting of classical algebraic topology), which is originally due to Tietze~\cite{tietze1908topologischen}. Modern and accessible definitions
of those can also be found in~\cite[Example~1.43]{hatcher2009algebraic} and
\cite[Chapter~V]{cohen2012course}. We suppose fixed a natural number~$m$:
although the notations do not mention it, the constructions below depend on this
natural number.

Given a natural number~$n$, we can view the sphere $\S{2n-1}$ as the subset
of~$\Cplx^n$ consisting of points whose euclidean norm is~$1$:
\[
  \S{2n-1}=\setof{(z_1,\ldots,z_n)\in\Cplx^n}{|z_1|^2+\ldots+|z_n|^2=1}
\]
Given $l_1,\ldots,l_n\in\N$ prime with $m$, we have a free action
\[
  \rho_{l_1,\ldots,l_n}:\Z_m\to\Aut(\S{2n-1})
\]
of~$\Z_m$ on $\S{2n-1}$ which rotates each $z_i$ by an angle of $2\pi l_i/m$:
the image of the generator $1\in\Z_m$ is
\[
  \rho_{l_1,\ldots,l_n}(1)(z_1,\ldots,z_n)
  =
  (\ce^{\frac{2\ci\pi l_1}m}z_1,\ldots,\ce^{\frac{2\ci\pi l_n}m}z_n)
\]
%
%
%
  %

We then define the \emph{lens space} $L(l_1,\ldots,l_n)$ as the (strict) quotient
$\S{2n-1}/\rho_{l_1,\ldots,l_n}$. The action $\rho_{l_1,\ldots,l_n}$ being free, the quotient map $\S{2n-1} \to L(l_1,\ldots,l_n)$ is thus covering, with $\Z_m$ as fiber~\cite[Proposition~1.40]{hatcher2009algebraic}, \ie we have a fiber sequence
\[
  \begin{tikzcd}
    \Z_m\ar[r,hook]&\S{2n-1}\ar[r,->>]&L(l_1,\ldots,l_n)
  \end{tikzcd}
\]
As such, it induces a long exact sequence in homotopy
groups~\cite[Theorem~4.41]{hatcher2009algebraic}:
\[
  \begin{tikzcd}[sep=small]
    \cdots\ar[r]&\pi_{k+1}(L)\ar[r]&\pi_k(\Z_m)\ar[r]&\pi_k(\S{2n-1})\ar[r]&\pi_k(L)\ar[r]&
    \cdots
  \end{tikzcd}
\]
where $L$ is a shorthand for $L(l_1,\ldots,l_n)$.
We have $\pi_0(\Z_m)=\Z_m$, $\pi_k(\Z_m)=0$ for $k>0$, and $\pi_k(\S{2n-1})=0$ for $k<2n-1$, from which we
deduce, using the above exact sequence, that $\pi_1(L)=\Z_m$ (when $n>1$) and $\pi_k(L)=0$ for
$0\leq k<2n-1$ with $k\neq 1$.

Moreover, given $l_{n+1}$ prime with $m$, there is a canonical morphism
from $L(l_1,\ldots,l_n)$ to $L(l_1,\ldots,l_n,l_{n+1})$, which adds $0$ as last
coordinate.
This allow us to define $L^\infty$, given an infinite sequence $l_1,l_2,...$ of integers prime with $m$, as the colimit of the diagram
\[
  \begin{tikzcd}
    L^0\ar[r]&L(l_1)\ar[r]&L(l_1,l_2)\ar[r]&\cdots
  \end{tikzcd}
\]
Since taking homotopy groups commutes with sequential colimits~\cite[Chapter 9, Section 4]{may1999concise}, we have
\[
  \pi_k(L^\infty)=
  \begin{cases}
    \Z_m&\text{if $k=1$,}\\
    0&\text{otherwise.}
  \end{cases}
\]
The space~$L^\infty$ is therefore a $K(\Z_m,1)$, or equivalently a $\B\Z_m$. As such, it does not depend on the choice of the parameters $l_i$, by unicity of the delooping, hence the notation. 
Lens spaces admit a CW-structure with one cell in every dimension $k\leq 2n-1$~\cite[Example~2.43]{hatcher2009algebraic}, 
thus $L^\infty$ inherits a CW structure with one cell in every dimension. The associated chain complex is
\[
  \begin{tikzcd}
    \cdots\ar[r,"0"]&\Z\ar[r,"\mu_m"]&\Z\ar[r,"0"]&\Z\ar[r,"\mu_m"]&\Z\ar[r,"0"]&\Z\ar[r]&0
  \end{tikzcd}
\]
where the map~$\mu_m$ denotes the multiplication by~$m$. Hence, the homology
groups $H_n(\Z_m)$ are: $\Z$ for $n=0$, $0$ for $n$ non-zero even, and $\Z_m$ for
$n$ odd.

\subsection{Definition in homotopy type theory}
\label{lens-type}
We now turn to the definition of lens spaces in type theory. Given an
integer~$l$, consider the group morphism
\[
  \phi^l:\Z\to\Z_m
\]
sending $1$ to~$l$. When~$l$ is prime to~$m$, we have a short exact sequence
\begin{equation}
  \label{ses-Zm}
  \begin{tikzcd}
    \Z\ar[r,hook,"\mu_m"]&\Z\ar[r,"\phi^l",->>]&\Z_m
  \end{tikzcd}
\end{equation}
By delooping, and \cref{delooping-short-exact-sequences}, we thus have a fiber
sequence
\begin{equation}
  \label{ses-BZm}
  \begin{tikzcd}
    \S1\ar[r,hook,"\B\mu_m"]&\S1\ar[r,->>,"\B \phi^l"]&\B\Z_m
  \end{tikzcd}
\end{equation}
(recall that $\B\Z=\S1$). By \cref{action-fiber-sequence}, we can view (3) as
an action of~$\Z_m$ on~$\S1$ (which turns the circle by a rotation of $2\pi/m$)
and whose homotopy quotient is~$\S1$. Therefore~$\S1$ is the first lens space.

By iterated join, see \cref{join-of-fiber-sequences}, we have, for integers
$l_1,\ldots,l_n$, and induced fiber sequence
\begin{equation*}
  \label{lens-space-fiber-sequence}
  \begin{tikzcd}[sep=15ex]
    (\S1)^{\join n}\ar[r,hook,"\B\mu_m^{\join n}"]&(\S1)^{\join_{\B\Z_m}n}\ar[r,->>,"\B\phi^{l_1}\join\ldots\join\B\phi^{l_n}"]&\B\Z_m
  \end{tikzcd}
\end{equation*}
Again, this can be viewed as an action of $\Z_m$ on the type $\S{2n-1}$ (which
is equal to~$(\S1)^{\join n}$), corresponding to the action
$\rho_{l_1,\ldots,l_n}$ of \cref{lens-topology}. By
\cref{action-fiber-sequence}, its homotopy quotient is the space
$(\S1)^{\join_{\B\Z_m}n}$. This justifies the following definition of lens
spaces.

\begin{definition}
  Given integers $l_1,\ldots,l_n$, relatively prime to~$m$, the associated
  \emph{lens space} is the type
  \[
    L(l_1,\ldots,l_n)\defd (\S1)^{\join_{\B\Z_m}n}
    \text.
  \]
\end{definition}

Suppose given an infinite sequence of integers $l_i$, with $i\in\N$, all
relatively prime to~$m$. For every~$n$ there is a canonical inclusion
\[
  L(l_1,\ldots,l_n)\to L(l_1,\ldots,l_n,l_{n+1})
\]
induced by the join and we write $L^\infty$ for the associated inductive
limit. By \cref{infinite-join}, we have:

\begin{theorem}
  \label{lens-deloop}
  The type $L^\infty$ is a delooping of~$\Z_m$.
\end{theorem}

\noindent
By uniqueness of deloopings\onappendix{ (\cref{delooping-unique})}, this type does not depend
on the choice of the sequence~$l_i$. Taking $l_i=1$ for every~$i$
is a reasonable canonical choice. However, as explained in the introduction, the study of lens spaces associated to various parameters is important in some applications other than delooping $\Z_m$~\cite{watkins}.

\subsection{The case of non-relatively prime parameters}
One can wonder what happens when we consider integer parameters $l_i$ which are
not supposed to be prime with~$m$. For~$l$ non-prime with~$m$, the sequence
\cref{ses-Zm} is not a fiber sequence, and therefore the sequence \cref{ses-BZm} is not
exact in general. In fact, the proper generalization is the following. Given a
natural number $k$ and a space $A$, we write $kA$ for the space
$A\sqcup\ldots\sqcup A$ with $k$ copies of~$A$. \onappendix{The following proposition is proved in \cref{non-prime-fiber-sequence-proof}.}

\begin{proposition}
  \label{non-prime-fiber-sequence}
  Given an arbitrary integer~$l$, with the notations of \cref{ses-BZm}, we
  have~$\fib{\B\phi^l}=k\S1$ with $k=\pgcd(l,m)$, \ie we have a fiber
  sequence
  \[
    \begin{tikzcd}[sep=large]
      k\S1\ar[r,hook,"k\B\mu_{m/k}"]&\S1\ar[r,->>,"\B \phi^l"]&\B\Z_m
    \end{tikzcd}
  \]
\end{proposition}

\begin{example}
  Consider situations where $m=5$ and $m=6$ respectively, and $l=2$ in both
  cases (thus $k=1$ and $k=2$, respectively). By (the proof in appendix of) \cref{non-prime-fiber-sequence},
  the kernel of $\B\phi^l:\S1\to\B\Z_m$ is the coequalizer of
  \[
    \begin{tikzcd}[sep=large]
      \Z_m\ar[r,shift left,"\id{}"]\ar[r,shift right,"\lambda i.(i+l)"']&\Z_m
    \end{tikzcd}
  \]
  For the two values of $m$, it can be pictured as
  
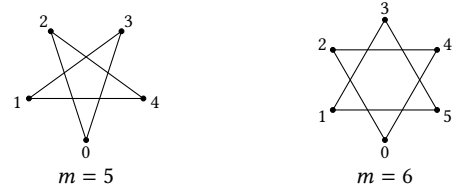
\begin{align*}
    \begin{tikzpicture}[baseline=(b.base),scale=.8]
      \coordinate (b) at (0,0);
      \foreach \i in {0,...,4} {
        \filldraw ({(\i+3/4)*360/5}:1) circle (.04);
        \draw ({((-\i+3)+3/4)*360/5}:1.2) node {$\scriptstyle\i$};
        \draw ({(\i+3/4)*360/5}:1) -- ({(\i+2+3/4)*360/5}:1);
      }
      \draw (0,-1.6) node {$m=5$};
    \end{tikzpicture}
    \qquad\qquad\qquad
    \begin{tikzpicture}[baseline=(b.base),scale=.8]
      \coordinate (b) at (0,0);
      \foreach \i in {0,...,5} {
        \filldraw ({(\i+1/2)*360/6}:1) circle (.04);
        \draw ({((-\i+4)+1/2)*360/6}:1.2) node {$\scriptstyle\i$};
        \draw ({(\i+1/2)*360/6}:1) -- ({(\i+2+1/2)*360/6}:1);
      }
      \draw (0,-1.6) node {$m=6$};
    \end{tikzpicture}
  \end{align*}
  and is respectively isomorphic to $\S1$ and $2\S1$.
\end{example}

Contrarily to the situation where~$l$ is prime with~$m$, the fiber sequence
of~\cref{non-prime-fiber-sequence} is not necessarily the delooping of an exact
sequence of groups. Namely, with $k\defd\pgcd(l,m)$, the type~$k\S1$ has $k$
connected components and is thus not the delooping of a group when
$k>1$. However, the computations still go on. Indeed, by iterated join, any sequence of integers $(l_1,\ldots,l_n)$ induces a fiber sequence
\[
  \begin{tikzcd}[sep=75]
    k_1\S1\join\ldots\join k_n\S1\ar[r,hook,"k_1\B\mu_{m/k_1}\join\ldots\join k_n\B\mu_{m/k_n}"]&(\S1)^{\join_{\B\Z_m}n}\ar[r,->>,"\B\phi^{l_1}\join\ldots\join\B\phi^{l_n}"]&[-42]\B\Z_m
  \end{tikzcd}
\]
which converges toward the sequence $1\into\B\Z_m\onto\B\Z_m$ (the second map is
an identity)
when~$n$ goes to the infinity, so that the spaces $(\S1)^{\join_{\B\Z_m}n}$
still play a role analogous to the one of lens spaces. However, an explicit
description of those is difficult to achieve contrarily to the case of lens spaces, see
\cref{cellularity}.

\subsection{A non-minimal resolution}
\label{non-minimal-resolution}
One can also wonder what would have happened if we had naively generalized the
construction for~$\B\Z_2$ as the infinite real projective space performed
in~\cite{buchholtz2017real} and recalled in \cref{RP}. We explain here that it
also gives rise to a model for $\B\Z_m$, although much larger than the one of
lens spaces.

\newcommand{\LL}[1]{\tilde L^{#1}}
The space $\B\Z_m$ is pointed and we write $f:1\to\B\Z_m$ for the pointed
map. The kernel $\ker f$ is the pullback of $f$ along itself, and thus
$\Loop\B\Z_m$ (see \cref{homotopy-limits}), hence $\Z_m$. We thus have a fiber
sequence
\[
  \begin{tikzcd}
    \Z_m\ar[r,hook]&1\ar[r,->>,"f"]&\B\Z_m
  \end{tikzcd}
\]
By iterated join on the map~$f$, one obtains a family of maps
\[
  f^{\join n}:\LL n\to\B\Z_m
\]
where the type $\LL n$, which is the source of this map, \ie the $n$-th iterated
join of $1$ relative to~$f$, converges to $\B\Z_m$ if we take the inductive
limit when $n$ goes to infinity, see \cref{join-of-fiber-sequences}.


While this does indeed produce a cellular delooping of~$\Z_m$, the resulting
model is much larger than the one of lens spaces. Already, for $n=2$, the
space~$\LL2$ we obtain is the pushout on the left, which can be pictured as on
the right:
\[
  \begin{tikzcd}
    \Z_m\ar[d]\ar[r]\ar[dr,phantom,pos=1,"\ulcorner"]&1\ar[d,dotted]\\
    1\ar[r,dotted]&\LL2
  \end{tikzcd}
  \qquad\qquad\qquad
  \begin{tikzcd}[sep=large]
    a
    \ar[-,r,bend left=60,pos=.52,"0"]
    \ar[-,r,bend left=20,"1"]
    \ar[phantom,r,bend left=-10,"\vdots"]
    \ar[-,r,bend left=-60,"m-1"']
    &b
  \end{tikzcd}
\]
We can see that it is much ``larger''
(in terms of number of cells) than the sphere~$\S1$, which is the first lens
space.
%
%
Similarly, higher spaces~$\LL n$ are much larger than the corresponding lens
spaces; in some sense, we are computing the free resolution of~$\Z_m$, which is
not at all minimal in terms of number of cells, see
\cite[Section~IV.5]{maclane2012homology}.
In contrast, lens spaces, which have one cell in every dimension, provide a minimal cellular resolution of~$\Z_m$: the homology of~$\Z_m$ (see \cref{lens-topology})
shows that we cannot build a delooping of~$\Z_m$ with less cells. Indeed this homology is non-trivial in degree $n$ for $n=0$ or $n$ odd, hence it implies that a given cellular model of~$\B\Z_m$ (whose cellular homology is the same as the homology of~$\Z_m$) should have \emph{at least} one cell in each dimension~$n$ for $n=0$ or $n$ odd. Moreover, the fact that these groups have torsion implies that the cellular model should also have \emph{at least} one cell in each dimension $n$ for $n$ even, otherwise the homology groups would be free since no relation in odd degree would be added. Hence a cellular model of~$\B\Z_m$ have \emph{at least} one cell in every dimension.
%


\section{Cellularity}
\label{cellularity}
We have already mentioned that the general constructions for deloopings of
groups (by HITs or torsors) do not easily allow eliminating to untruncated types.
%
%
We show here that our alternative model of~$\B\Z_m$, provided by infinite lens
spaces constructed above, admits an inductive description.
This constitutes a type-theoretic counterpart of the CW-complex structure of topological lens
spaces, and induce an induction principle which allows eliminating into
general types.



We begin by recalling the following classical flattening lemma for pushouts,
see~\cite[Lemma~2.2.5]{rijke2018classifying} for a detailed proof.

\begin{lemma}[Flattening for pushouts]
  \label{flattening-pushout}
  Consider a pushout square
  \[
    \begin{tikzcd}
      X\ar[d,"f"']\ar[r,"g"]\ar[dr,phantom,pos=1,"\ulcorner"]&B\ar[d,"j"]\\
      A\ar[r,"i"']&A\sqcup_XB
    \end{tikzcd}
  \]
  with $p:i\circ f=j\circ g$ witnessing for its commutativity, together with a
  type family $P:A\sqcup_XB\to\U$. Then the following square of total spaces is
  also a pushout
  \[
    \begin{tikzcd}
      \Sigma X.(P\circ i\circ f)\ar[d,"\Sigma f.(\lambda\_.\id{})"']\ar[r,"\Sigma g.e"]&\Sigma B.(P\circ j)\ar[d,"\Sigma j.(\lambda\_.\id{})"]\\
      \Sigma A.(P\circ i)\ar[r,"\Sigma i.(\lambda\_.\id{})"']&\Sigma(A\sqcup_XB).P
    \end{tikzcd}
  \]
  where $e:(x:X)\to P(i(f(x)))\to P(j(g(x)))$ is the canonical morphism induced
  by~$p$, \ie $e(x)\defd\subst P{\happly px}$.
\end{lemma}

\noindent
Given a pointed type~$A$, the space $\Sigma(x:A).(\pt{}=x)$ is always
contractible. The following result is a dependent generalization of this fact, and is used later on:

\begin{lemma}
  \label{dependent-singleton}
  Given a type~$X$, a family $P:X\to\U$, elements $a,b:X$, a path $p:a=b$, and
  an element $x:P(a)$, the type
  $
    \Sigma(y:P(b)).(x=^P_py)
  $
  is contractible.
\end{lemma}

\begin{proof}
  By path induction, it is enough to show the contractibility of
  \[
    \Sigma(y:P(a)).(x=^P_{\refl{}}y)
  \]
  \ie of the type
  \[
    \Sigma(y:P(a)).(x=y)
  \]
  which is known to be true~\cite[Lemma 3.11.8]{hottbook}.
\end{proof}

\subsection{Sigma types preserve joins}
We can now show that sigma types preserve joins relative to morphisms. Under the
equivalence of \cref{fibrations-families} between fibrations and type families,
this generalizes the fact that fibers preserve joins,
already encountered in~\cref{fib-join}.

\begin{theorem}
  \label{sigma-join}
  Suppose given maps $f:A\to X$ and $g:B\to X$, a type family $P:X\to\U$. Then
  we have
  \[
    \Sigma(A\join_XB).\pa{P\circ(f\join g)}
    =
    \pa{\Sigma A.\pa{P\circ f}}
    \join_{\Sigma X.P}
    \pa{\Sigma B.\pa{P\circ g}}
    \text.
  \]
  where $A\join_XB$ is the source of the join of~$f$ and $g$, and the type on
  the right is the source of the join of the maps
  \begin{align*}
    \Sigma f.(\lambda\_.\id{}):\Sigma A.(P\circ f)&\to\Sigma X.P
    \\
    \Sigma g.(\lambda\_.\id{}):\Sigma B.(P\circ g)&\to\Sigma X.P
  \end{align*}
\end{theorem}
\begin{proof}
  By definition of the join of~$f$ and~$g$, we have a diagram
  \[
    \begin{tikzcd}
      A\times_X B\ar[d,"\pi_1"']\ar[r,"\pi_2"]&B\ar[d,"\sndinj"]\ar[ddr,bend left,"g"]\\
      A\ar[drr,bend right=20,"f"']\ar[r,"\fstinj"']&A\join_XB\ar[dr,"f\join g"description]\\
      &&X
    \end{tikzcd}
  \]
  where the outer square is a pullback and the inner square is a pushout. We denote by
  \begin{align*}
    p_1:\fstinj\circ(f\join g)&=f&
    p_2:\sndinj\circ(f\join g)&=g&
    p:f\circ\pi_1&=g\circ\pi_2
  \end{align*}
  the equalities witnessing for commutativity of various subdiagrams. By the flattening lemma for
  pushouts (\cref{flattening-pushout}) applied to the type family
  \[
    P\circ(f\join g):A\join_XB\to\U
  \]
  we have a pushout
  \[
    \begin{tikzcd}
      \Sigma(A\times_XB).P\circ(f\join g)\circ\fstinj\circ\pi_1\ar[d,"\Sigma\pi_1.(\lambda\_.\id{})"']\ar[r,"\Sigma\pi_2.e"]&\Sigma B.P\circ(f\join g)\circ\sndinj\ar[d,"\Sigma\sndinj.(\lambda\_.\id{})"]\\
      \Sigma A.P\circ(f\join g)\circ\fstinj\ar[r,"\Sigma\fstinj.(\lambda\_.\id{})"']&\Sigma(A\join_XB).P\circ(f\join g)
    \end{tikzcd}
  \]
  where $e$ is the canonical function induced by transport along the canonical
  composite equality
  \[
    (f\join g)\circ\fstinj\circ\pi_1
    =
    f\circ\pi_1
    =
    g\circ\pi_2
    =
    (f\join g)\circ\sndinj\circ\pi_2
  \]
  induced by the above ones. By using the identities $p_1$, $p_2$ and~$p$, this
  simplifies as
  \begin{equation}
    \label{eq:sigma-join-pushout1}
    \begin{tikzcd}[column sep=50]
      \Sigma(A\times_XB).P\circ f\circ\pi_1\ar[d,"\Sigma\pi_1.(\lambda\_.\id{})"']\ar[r,"\Sigma\pi_2.(\lambda x.\subst P{p(x)})"]&\Sigma B.P\circ g\ar[d,"\Sigma\sndinj.\substinv P{p_2}"]\\
      \Sigma A.P\circ f\ar[r,"\Sigma\fstinj.\substinv P{p_1}"']&\Sigma(A\join_XB).P\circ(f\join g)
    \end{tikzcd}
  \end{equation}
  In order to conclude, it is enough to show that the upper-left span coincides
  with the upper-left span of the following pullback diagram:
  \begin{equation}
    \label{eq:sigma-join-pushout2}
    \begin{tikzcd}[column sep=small]
      (\Sigma A.(P\circ f))\times_{\Sigma X.P}(\Sigma B.(P\circ g))\ar[d,"\pi_1"']\ar[r,"\pi_2"]&\Sigma B.(P\circ g)\ar[ddr,bend left,"\Sigma g.(\lambda\_.\id{})"description]\\
      \Sigma A.(P\circ f)\ar[drr,bend right=15,"\Sigma f.(\lambda\_.\id{})"description]\\
      &&\Sigma X.P
    \end{tikzcd}
  \end{equation}
  Namely, the pushout of the upper-left span of \cref{eq:sigma-join-pushout2}
  is, by definition of the join,
  \[
    (\Sigma A.(P\circ f))\join_{\Sigma X.P}(\Sigma B.(P\circ g))
  \]
  and it will coincide with the pushout of the upper-left span of
  \cref{eq:sigma-join-pushout1}, which is
  \[
    \Sigma(A\join_XB).P\circ(f\join g)
    \text.
  \]
  We show that the upper-left objects coincide (the fact that the map coincide
  can be shown by transporting along the resulting equality, which is left to
  the reader). We namely have
  \begin{align*}
    (\Sigma A.&(P\circ f))\times_{\Sigma X.P}(\Sigma B.(P\circ g))
    \\
    &=\Sigma((a,x):\Sigma A.(P\circ f)).\Sigma((b,y):\Sigma B.(P\circ g)).\\
    &\qquad\quad(\Sigma f.(\lambda\_.\id{}))(a,x)=(\Sigma g.(\lambda\_.\id{}))(b,y)
    \\
    &=\Sigma(a:A).\Sigma(x:P(f(a))).\Sigma(b:B).\Sigma(y:P(g(b))).\\
    &\qquad\quad\Sigma(p:f(a)=g(b)).(x=^P_py)
    \\
    &=\Sigma(a:A).\Sigma(b:B).\Sigma(p:f(a)=g(b)).\Sigma(x:P(f(a))).\\
    &\qquad\quad\Sigma(y:P(g(b))).(x=^P_py)
    \\
    &=\Sigma(a:A).\Sigma(b:B).\Sigma(p:f(a)=g(b)).\Sigma(x:P(f(a))).1
    \\
    &=\Sigma(a:A).\Sigma(b:B).(f(a)=g(b))\times P(f(a))
    \\
    &=\Sigma(A\times_XB).(P\circ f\circ\pi_1)
  \end{align*}
  Most of the steps of this reasoning can easily be justified by associativity
  and commutativity properties of $\Sigma$-types. The only step which is not of this
  nature, in the middle, uses the fact that the type
  \[
    \Sigma(y:P(g(b))).(x=^P_py)
  \]
  is contractible for $a:A$, $b:B$, $p:f(a)=g(b)$ and $x:P(f(a))$, which follows
  from \cref{dependent-singleton}.
  %
  %
\end{proof}

\noindent
With the notations of the above theorem, we have the following interesting
particular case. When~$\Sigma X.P$ is contractible (\ie when~$P$ is the
universal cover of~$X$), we have
\[
  \Sigma(A\join_XB).\pa{P\circ(f\join g)}
  =
  \pa{\Sigma A.\pa{P\circ f}}
  \join
  \pa{\Sigma B.\pa{P\circ g}}
\]
This is essentially the result which is used in
\cite[Theorem~III.4]{buchholtz2017real}.

Under the fibration-family correspondence of \cref{fibrations-families}, the
above theorem can be reformulated as follows:

\begin{theorem}
  \label{sigma-join-descent}
  Suppose given maps~$f:A\to X$ and $g:B\to X$ and a map $h:Y\to X$, we have
  \[
    (A\join_XB)\times_XY=(A\times_XY)\join_Y(B\times_XY)
  \]
  where the pullbacks are taken over the expected maps. This can be summarized
  by the diagram
  \[
    \begin{tikzcd}[sep=small]
      &(A\times_XY)\times_A(B\times_XY)\ar[dd]\ar[dl]\ar[rr]&[-40pt]&B\times_XY\ar[dl]\ar[dd]\ar[ddr,bend left]\\
      A\times_XY\ar[drrrr,bend right=40]\ar[dd]\ar[rr]&&(A\join_XB)\times_XY\ar[dd]\ar[drr,bend left=7]\\
      &A\times_XB\ar[dl]\ar[rr]&&B\ar[ddr,bend left,pos=0.6,"g"description]\ar[dl]&Y\ar[dd]\\
      A\ar[drrrr,bend right=10,"f"description]\ar[rr]&&A\join_XB\ar[drr,"f\join g"description]\\
      &&&&X
    \end{tikzcd}
  \]
  where the bottom diagram is the pushout of pullback defining the join, the
  vertical squares are pullbacks
  and the theorem states that the top square is a pushout.
\end{theorem}


\noindent
\Cref{sigma-join-descent}, which will not be needed elsewhere, is a direct application of descent.
As a particular case, when $Y$ is a point (or, more generally, contractible), the theorem simplifies as
\[
  \fib{f\join g}(x)=\fib{f}(x)\join\fib{g}(x)
\]
where $x$ is the image by~$h$ of the point of~$Y$, and we recover
\cref{fib-join}.

\subsection{A characterization of relations}
We have seen at the end of \cref{lens-type} that the construction of lens spaces
does not depend on the choice of parameters $l_i$, so that we can always
take~$l_i=1$. In order to simplify notations, we thus write
\[
  L^n\qdefd L(1,\ldots,1)
\]
(with $n$ occurrences of~$1$ on the right) and will only work with such lens
spaces in the following (but all constructions would generalize in the expected
way to varying parameters $l_i$). We also write $\phi:\Z\to\Z_m$ instead
of~$\phi^1$, so that $L^0=0$ and $L^{n+1}$ is obtained as the join
\[
  L^{n+1}\qdefd L^n\join_{\B\Z_m}\S1
\]
\ie as a pushout of the following pullback:
\begin{equation}
  \label{Ln-definition}
  \begin{tikzcd}[row sep=small,column sep=60]
    L^n\times_{\B\Z_m}\S1\ar[d]\ar[r]&\S1\ar[d]\ar[ddr,bend left=15,"\B\phi"description]\\
    L^n\ar[drr,bend right=10,"(\B\phi)^{\join n}"description]\ar[r]&L^{n+1}\ar[dr,"(\B\phi)^{\join(n+1)}"description]\\
    &&\B\Z_m\\
  \end{tikzcd}
\end{equation}
Our goal in this section is to provide a concrete description of the upper left
space, which will play a central role in our recursion principle for $\B\Z_m$.

We write
\[
  R
  \qdefd
  \Sigma(x:\S1).\Sigma(y:\S1).(\B\phi(x)=\B\phi(y))
\]
for the pullback $\S1\times_{\B\Z_m}\S1$, \ie
\[
  \begin{tikzcd}
    R\ar[d,dotted]\ar[r,dotted]\ar[dr,phantom,pos=0,"\lrcorner"]&\S1\ar[d,"\B\phi"]\\
    \S1\ar[r,"\B\phi"']&\B\Z_m
  \end{tikzcd}
\]
In order to characterize~$R$, we first show that it is a delooping (of its loop space).

\begin{lemma}
  The type $R$ is a pointed connected groupoid.
\end{lemma}
\begin{proof}
  The distinguished elements of~$\S1$ and $\B\Z_m$ induce a distinguished
  element
  \[
    (\pt{\S1},\pt{\S1},\refl{\pt{\B\Z_m}})
  \]
  in~$R$, making it pointed.
  Also the type $R$ is a groupoid as a pullback of
  groupoids~\cite[Theorem~7.1.8]{hottbook}.
  We are left with showing that it is connected. Suppose given a point
  $(x,y,r)$ of~$R$: we want to show that there merely exists a path from the
  distinguished point of~$R$ to it. Since this is a proposition and $\S1$ is
  connected, we can suppose given paths $\pt{}=x$ and $\pt{}=y$ in~$\S1$, and,
  by path induction, those paths can be supposed to be $\refl{}$. We are thus
  left with proving $(\pt{},\pt{},\refl{})=(\pt{},\pt{},r)$ for an arbitrary
  path $r:\Loop\B\Z_m$. By the characterization of paths in
  $\Sigma$-types~\cite[Theorem~2.7.2]{hottbook} and transport along identity
  types~\cite[Theorem~2.11.3]{hottbook}, this amounts to find paths
  $p,q:\Loop\S1$ together with a proof that
  \[
    r
    =
    \sym{\ap{(\B\phi)}p}\pcomp\refl{}\pcomp\ap{(\B\phi)}q
  \]
  Since $\Loop\S1=\Z$ and $\Loop\B\Z_m=\Z_m$, this amounts to show that the
  canonical morphism
  $\Z\times\Z\to\Z_m$ sending $(i,j)$ to $j-i$
  is surjective, which is the case: it admits the map $i\mapsto(0,i)$ as a
  section.
\end{proof}

\begin{remark}
  The property of being connected is not preserved in general under
  pullbacks.
  For instance, the pullback of the map $1\to\S1$ along itself is $\Loop\S1$,
  \ie $\Z$, which is not connected. The proof of the connectedness of~$R$ thus
  had to be specific.
\end{remark}

\noindent
By \cref{groups-vs-pointed-connected-groupoids}, we thus have $R=\B\Loop R$, \ie
$R$ is determined by its loop space, which is a group that can be computed in
the following way.

\begin{lemma}
  The type $\Loop R$ is obtained as the following pullback:
  \[
    \begin{tikzcd}
      \Loop R\ar[d,dotted]\ar[r,dotted]\ar[dr,phantom,pos=0,"\lrcorner"]&\Z\ar[d,"q"]\\
      \Z\ar[r,"q"']&\Z_m
    \end{tikzcd}
  \]
  where~$q$ is the canonical quotient map. Hence $R$ is a delooping of the following subgroup of $\Z^2$
  \[
      \setof{(j,k)\in\Z\times\Z}{k-j=0\mod m}\text{.}
  \]
\end{lemma}
\begin{proof}
  Given a pointed type~$A$, the loop space functor can be described as a covariant hom functor (we have $\Loop A=\S1\pto A$, see \cite[Lemma 6.2.9]{hottbook}), so it preserves limits. The loop space of~$R$ is thus
  \begin{align*}
    \Loop R
    &=\Loop (\S1\times_{\B\Z_m}\S1)&&\text{by definition of~$R$}\\
    &=\Loop\S1\times_{\Loop\B\Z_m}\Loop\S1&&\text{by commutation of $\Omega$ with pullbacks}\\
    &=\Z\times_{\Z_m}\Z
  \end{align*}
  From which we conclude.
\end{proof}

\noindent
An even more explicit description of this group is:

\begin{lemma}
  $\Loop R=\Z^2$.
\end{lemma}
\begin{proof}
  It is easily shown that $(1,1)$ and $(0,m)$ form a basis of $\Loop R$, which is thus isomorphic to~$\Z^2$.
\end{proof}

\begin{corollary}
  The type~$R$ is isomorphic to the torus.

\end{corollary}
\begin{proof}
  The torus is also a delooping of~$\Z^2$, so we conclude by unicity of the
  delooping of a given group.
\end{proof}

\noindent
Note that once we fix a basis of~$\Loop R$ (for instance $\{(1,1),(0,m)\}$),
we have an explicit isomorphism between $R$ and the torus given by the delooping of the associated
``base change'' morphism $\Z^2\to\Loop R$.

We write $R^{\join n}$ for the iterated product of~$n$ instances of the second projection $\snd:R\to\S1$:
\[
  R^{\join n}
  \qdefd
  R\join_{\S1}R\join_{\S1}\ldots\join_{\S1}R
\]
(note that this is a dependent join, we should write $R^{\join_{\S1}n}$ but this is ugly). As a consequence of \cref{sigma-join}, we have the following theorem.

\begin{theorem}
  \label{lens-pushout}
  We have, for every~$n\in\N$,
  \[
    L^n\times_{\B\Z_m}\S1=R^{\join n}
  \]
  \ie the upper-left square of \cref{Ln-definition} is the pushout
  \[
    \begin{tikzcd}
      R^{\join n}\ar[d]\ar[r]\ar[dr,phantom,"\ulcorner",pos=1]&\S1\ar[d]\\
      L^n\ar[r]&L^{n+1}
    \end{tikzcd}
  \]
\end{theorem}
\begin{proof}
  For the base case, we have $L^0=0$ and the result is immediate.
  For the inductive case, we have
  \begin{align*}
    L^{n+1}&\times_{\B\Z_m}\S1
    \\
    &=\Sigma(x:L^{n+1}).\Sigma(y:\S1).(\B\phi^{\join(n+1)}(x)=\B\phi(y))
    \\
    &=\Sigma(x:(L^n\join_{\B\Z_m}\S1)).\Sigma(y:\S1).(\B\phi^{\join(n+1)}(x)=\B\phi(y))
    \\
    &=\pa{\Sigma(x:L^n).\Sigma(y:\S1).(\B\phi^{\join n}(x)=\B\phi(y))}\join_{\Sigma(x:\B\Z_m).P(x)}\\
    &\phantom{\ =\ }\pa{\Sigma(x:\S1).\Sigma(y:\S1).(\B\phi(x)=\B\phi(y))}
    \\
    &=(L^n\times_{\B\Z_m}\S1)\join_{\S1} R
    \\
    &=R^{\join n}\join_{\S1}R
    \\
    &=R^{\join(n+1)}
  \end{align*}
  Above, in order to go from the second to the third line, we apply
  \cref{sigma-join} with $A=L^n$, $B=\S1$, $X=\B\Z_m$, $f=(\B\phi)^{\join n}$,
  $g=\B\phi$ and $P:\B\Z_m\to\U$ defined by
  $
  P(x)\defd\Sigma(y:\S1).(x=\B\phi(y))
  $
  which satisfies
  \begin{align*}
    \Sigma(x:\B\Z_m).P(x)
    &=
    \Sigma(x:\B\Z_m).\Sigma(y:\S1).(x=\B\phi(y))
    \\
    &=
    \Sigma(y:\S1).\Sigma(x:\B\Z_m).(x=\B\phi(y))
    \\
    &=
    \Sigma(y:\S1).1
    \\
    &=
    \S1    
  \end{align*}
  This allows us to conclude.
\end{proof}

\subsection{A recursion principle}
Fix an arbitrary type~$X$. As a direct consequence of the above theorem, maps
$L^{n+1}\to X$ correspond to triples consisting of a map $L^n\to X$, a map
$\S1\to X$ and an equality witnessing the commutation of the square
\[
  \begin{tikzcd}
      R^{\join n}\ar[d]\ar[r]\ar[dr,phantom,"\ulcorner",pos=1]&\S1\ar[d]\\
      L^n\ar[r]&X
  \end{tikzcd}
\]
When $X$ is an $n$-type, we have that maps $\B\Z_m\to X$ correspond to maps
$L^n\to X$, which can be defined as above, by induction on~$n$. For general
types~$X$, we have $\B\Z_m=\colim_n L^n$ and thus, since hom functors commutes with
colimits, we have the following result, which allows computing maps out of the
delooping of~$\Z_m$.

\begin{theorem}
  \label{cyclic-recursion}
  Given a type~$X$, the maps $\B\Z_m\to X$ are the limits of maps $L^n\to X$.
\end{theorem}

\subsection{Applications}
\label{cohomology}
Let us mention some possible applications of this result. Given a group~$G$, its
$n$-th integral \emph{cohomology group} is the space
$H_n(G)\defd\strunc{\B G\to K(\Z,n)}$
of homotopy classes of maps from $\B G$ to the space $K(\Z,n)$, which is an $n$-type~\cite{licata2014eilenberg,buchholtz2020cellular}. \Cref{cyclic-recursion} should thus allow to perform explicit computations of cohomology classes of~$\Z_m$ of degree $n$, even when $n>1$.

As another application, the above theorem allows defining maps from $\B\Z_m$ to
$n$-types or types (whereas the HIT definition of \cite{licata2014eilenberg}
only allows eliminating to sets, the case $n=0$), thus enabling us to define actions of~$\Z_m$
on higher homotopy types. Indeed, recall that maps $f : \B G \to \U$ with $f(\pt{}) = X$ encodes actions of $G$ on the type $X$, thus being able to eliminate to universes of ($n$-)types provide a way to define actions on ($n$-)types.

\subsection{Cellularity}
\Cref{lens-pushout} provides an inductive description of lens spaces as iterated pushouts, and we have seen above that this description can be useful in order to perform inductive computations on lens spaces.
However, in order for some of the traditional techniques to be directly applicable, such as the computation of cohomology groups~\cite{buchholtz2020cellular}, we should show that they are \emph{cellular}, \ie that they can be obtained by pushouts of a particular form, corresponding to attaching disks along boundary they boundaries. We claim that this is the case for the lens spaces we define here, but leave the rather involved computations for future work.

We can however show here that the types $\LL n$ of the ``non-minimal resolution''  of \cref{non-minimal-resolution} are cellular. Namely, by a direct generalization of the arguments of \cref{RP} (from~$2$ to~$m$), for any $n\in\N$, we have $\LL n\times_{\B{\Z_m}}1=\Z_m^{\join(n+1)}$ and we thus have a pushout
\begin{equation}
  \label{LL-pushout}
  \begin{tikzcd}
    \Z_m^{\join(n+1)}\ar[d]\ar[r]\ar[dr,phantom,pos=1,"\ulcorner"]&1\ar[d]\\
    \LL n\ar[r]&\LL{n+1}
  \end{tikzcd}
\end{equation}
Moreover, we have the following reformulation of the join product~$\Z_m^{\join(n+1)}$.
We write $A\vee B$ for the wedge sum of two pointed spaces, which is the pushout of $A\sqcup_1B$ of the pointing maps $1\to A$ and $1\to B$, and more generally we write $\bigvee_nA$ for the wedge sum of $n$ instances of~$A$. It is not difficult to show that join distributes over wedge: for types $A$, $B$ and $C$, we have
\begin{equation}
  \label{join-wedge}
  (A\vee B)\join C
  =
  (A\join C)\vee(A\join B)
\end{equation}

\begin{lemma}
  For any $n\in\N$, we have $\Z_m^{\join n}=\bigvee_{(m-1)^n}\S{n-1}$.
\end{lemma}
\begin{proof}
  We reason by induction on~$n$. For the base case $n=0$, we have $0=\bigvee_{1}\S{-1}$ and, by induction, we have
  \begin{align*}
    \Z_m^{\join(n+1)}
    &=\Z_m^{\join n}\join\Z_m
    \\
    &=\Bigpa{\bigvee_{(m-1)^n}\S{n-1}}\join\Z_m&&\text{by induction}
    \\
    &=\bigvee_{(m-1)^n}\pa{\S{n-1}\join\Z_m}&&\text{by \cref{join-wedge}}
    \\
    &=\bigvee_{(m-1)^n}\susp^n\Z_m&&\text{because $\S{n-1}=(\S0)^{\join n}$}
    \\
    &=\bigvee_{(m-1)^n}\susp^{n-1}\Bigpa{\bigvee_{m-1}\S1}&&\text{because $\Sigma\Z_m=\bigvee_{m-1}\S1$}
    \\
    &=\bigvee_{(m-1)^n}\bigvee_{m-1}\susp^{n-1}\S1&&\text{by \cref{join-wedge}}
    \\
    &=\bigvee_{(m-1)^{n+1}}\S n
  \end{align*}
  from which we conclude.
\end{proof}

\noindent
%
The pushout \cref{LL-pushout} thus shows that $\LL{n+1}$ is cellular since it is obtained from~$\LL n$ by attaching $(m-1)^{n+1}$ disks along $n$-spheres.
Namely, the above properties imply that we have a pushout of the form
\[
  \begin{tikzcd}
    \Fin{(m-1)^{n+1}}\times\S{n}\ar[d]\ar[r,"\fst"]\ar[dr,phantom,pos=1,"\ulcorner"]&\Fin{(m-1)^{n+1}}\ar[d]\\
    \LL n\ar[r]&\LL{n+1}
  \end{tikzcd}
\]
where $\Fin k$ denotes a set with $k$ elements.



\section{Delooping dihedral groups}
\label{delooping-dihedral}
In this section, we explain how to construct the delooping of a group~$P$ which
admits a group~$G$ as (strict) quotient for which we already know a
delooping. This situation is very general: in particular, it includes the case
where~$P$ is the semidirect product of~$G$ with another group~$H$. As an
illustration, we explain how to construct deloopings for dihedral groups.



\subsection{Delooping short exact sequences}
\label{internal-semidirect-product}
Recall from \cref{delooping-short-exact-sequences} that any short exact sequence
of groups
\begin{equation}
  \label{sdp-exact}
  \begin{tikzcd}
    1\ar[r]&H\ar[r]&P\ar[r,"f"]&G\ar[r]&1
  \end{tikzcd}
\end{equation}
induces, by delooping, a fiber sequence
\[
  \begin{tikzcd}
    \B H\ar[r,hook]&\B P\ar[r,->>,"\B f"]&\B G
  \end{tikzcd}
\]
(note that this includes in particular the case where~$P$ is the semidirect
product of~$G$ and~$H$, \ie when the short exact sequence is split).
As explained in \cref{fibrations-families}, in such a situation, we have an
action of $G$ on~$\B H$ given by $\fib f:\B G\to\U$, whose homotopy quotient
$\Sigma(x:\B G).\fib f(x)$ is $\B P$. This thus gives us a way to construct a
delooping of~$P$ from a delooping of~$G$, as illustrated below.

\subsection{Dihedral groups}
Given a natural number~$m$, we write $D_m$ for $m$-th \emph{dihedral group},
which is the group of symmetries of a regular polygon with~$m$ sides. This group
is generated by a rotation~$r$ and a symmetry~$s$, and admits the presentation
\[
  D_m
  \qdefd
  \pres{r,s}{r^m=1,s^2=1,rs=sr^{m-1}}
\]
From this presentation, it can be seen that we have
a short exact sequence
\begin{equation}
  \label{dihedral-exact-sequence}
  \begin{tikzcd}
    \Z_m\ar[r,"\kappa"]&D_m\ar[r,"\pi"]&\Z_2
  \end{tikzcd}
\end{equation}
where the maps are characterized by $\kappa(1)=r$, $\pi(r)=0$ and $\pi(s)=1$,
see for instance~\cite[Chapter~XII]{mac1999algebra}.

By the reasoning of \cref{internal-semidirect-product}, we have the following
delooping for~$D_m$:
\[
  \B D_m
  \quad\defd\quad
  \Sigma(x:\B\Z_2).\fib\pi(x)
\]
By curryfication, we thus have a recursion principle for~$\B D_m$. Namely, given
a type~$X$, we have
\[
  \B D_m\to X
  \qquad=\qquad
  (x:\B\Z_2)\to\fib\pi(x)\to X
\]
Here, the arrow on the right can be computed by eliminating from~$\B\Z_2$ as
explained in \cref{cyclic-recursion}.

In order to be fully satisfied, we still need a usable description
of~$\fib\pi(x)$. This can be achieved as follows.
The short exact sequence~\cref{dihedral-exact-sequence} is split, with the
section $\sigma:\Z_2\to D_n$ such that $\sigma(1)=s$ (the existence of such a
splitting corresponds to the fact that $D_m$ is a semidirect product of $\Z_m$
and~$\Z_2$). This splitting induces an action $\phi:\Z_2\to\Aut(\Z_m)$ which is
the group morphism determined by $\phi(1)(k)=-k$.
Moreover, writing $\Upt[\B\Z_m]$ for the type $\Upt$ of pointed types, pointed at $\B\Z_m$, we have $\Aut(\Z_m)=\Loop\Upt[\B\Z_m]$, thus $\phi$ can be seen as an action of~$G$ on~$\B\Z_m$.
Its delooping can be shown to coincide with~$\fib\pi$, so that the action given by the fiber sequence induced by \cref{dihedral-exact-sequence} is the delooping of the
action induced by the splitting. The point here is that, up to the correspondence between actions on~$\Z_m$ and actions on~$\B \Z_m$, $\B\phi$ is an action on $\Z_m$ (a set), so it can be explicitely defined by using the Finster-Licata
construction of $\B \Z_2$~\cite{licata2014eilenberg}. Indeed, Set is a groupoid, so we can define $\B \phi : \B\Z_2 \to\HSet $ by giving its value on generators and relations. Hence, we have obtained a description of~$\fib\pi$ which is relevant from a computational point of view, allowing us to make use of the recursion principle for $\B D_m$.
This will be detailed in another paper, along with a general study of delooping
of semidirect products.

We should finally mention that the above construction can directly be
generalized in order to construct deloopings for all (split) \emph{metacyclic
  groups}: those are groups~$P$ for which there is a short exact sequence
\cref{sdp-exact} with $G$ and $H$ cyclic, and include dihedral, quasidihedral
and dicyclic groups as particular cases.



\section{Conclusion and future work}
\label{conclusion}
We believe that the general recursion principles we gave for $\B\Z_m$ in
\cref{cellularity} and for $\B D_m$ in \cref{delooping-dihedral} will be useful
to perform computations with cyclic and dihedral groups involving
non-trivial higher geometry: for instance, computing cohomology classes or
defining higher actions of these groups, as explained in \cref{cohomology}. This
would be a great step forward in synthetic group theory since homotopy type
theory, as opposed to plain type theory, is mainly about reasoning with objects
which are non-trivial from a homotopical perspective. This is left for future
work.

We also plan to use other geometric models in order to obtain cellular models of
delooping for other classical groups, while defining in the process some
important spaces of algebraic topology which are not formalized yet. In
particular, we are working on the construction in homotopy type theory of the
hypercubical manifold (a space studied by Poincaré during his quest to define
homology spheres~\cite{poincare1895analysis}), and higher variants of this space
using the join construction and the action-as-fibration paradigm.

As mentioned at the end of \cref{delooping-dihedral}, we will further
investigate the delooping of semidirect products of groups. This should provide
us with a modular way of constructing cellular models of deloopings from already
known ones.

Finally, we plan to have an Agda formalization of our work (the implementation
of lens spaces in homotopy type theory and the recursion principle of $\B\Z_m$
associated). This would first require developing basic constructions such as the
join of maps, which are unfortunately not present at the moment in the standard
cubical library.




\bibliographystyle{ACM-Reference-Format}
\bibliography{papers}


\begin{thebibliography}{27}


\ifx \showCODEN    \undefined \def \showCODEN     #1{\unskip}     \fi
\ifx \showDOI      \undefined \def \showDOI       #1{#1}\fi
\ifx \showISBNx    \undefined \def \showISBNx     #1{\unskip}     \fi
\ifx \showISBNxiii \undefined \def \showISBNxiii  #1{\unskip}     \fi
\ifx \showISSN     \undefined \def \showISSN      #1{\unskip}     \fi
\ifx \showLCCN     \undefined \def \showLCCN      #1{\unskip}     \fi
\ifx \shownote     \undefined \def \shownote      #1{#1}          \fi
\ifx \showarticletitle \undefined \def \showarticletitle #1{#1}   \fi
\ifx \showURL      \undefined \def \showURL       {\relax}        \fi
\providecommand\bibfield[2]{#2}
\providecommand\bibinfo[2]{#2}
\providecommand\natexlab[1]{#1}
\providecommand\showeprint[2][]{arXiv:#2}

\bibitem[Awodey and Warren(2009)]%
        {awodey2009homotopy}
\bibfield{author}{\bibinfo{person}{Steve Awodey} {and}
  \bibinfo{person}{Michael~A Warren}.} \bibinfo{year}{2009}\natexlab{}.
\newblock \showarticletitle{Homotopy theoretic models of identity types}. In
  \bibinfo{booktitle}{\emph{Mathematical proceedings of the cambridge
  philosophical society}}, Vol.~\bibinfo{volume}{146}. Cambridge University
  Press, \bibinfo{pages}{45--55}.
\newblock
\urldef\tempurl%
\url{https://doi.org/10.1017/S0305004108001783}
\showDOI{\tempurl}
\showeprint{0709.0248}


\bibitem[Bezem et~al\mbox{.}({[n.\,d.]})]%
        {symmetry}
\bibfield{author}{\bibinfo{person}{Marc Bezem}, \bibinfo{person}{Ulrik
  Buchholtz}, \bibinfo{person}{Pierre Cagne}, \bibinfo{person}{Bjørn~Ian
  Dundas}, {and} \bibinfo{person}{Daniel~R. Grayson}.}
  \bibinfo{year}{[n.\,d.]}\natexlab{}.
\newblock \bibinfo{title}{Symmetry}.
\newblock
  \bibinfo{howpublished}{\url{https://github.com/UniMath/SymmetryBook}}.
\newblock


\bibitem[Brunerie(2016)]%
        {brunerie2016homotopy}
\bibfield{author}{\bibinfo{person}{Guillaume Brunerie}.}
  \bibinfo{year}{2016}\natexlab{}.
\newblock \emph{\bibinfo{title}{On the homotopy groups of spheres in homotopy
  type theory}}.
\newblock \bibinfo{thesistype}{Ph.\,D. Dissertation}.
\newblock
\showeprint{1606.05916}


\bibitem[Brunerie(2019)]%
        {brunerie2019james}
\bibfield{author}{\bibinfo{person}{Guillaume Brunerie}.}
  \bibinfo{year}{2019}\natexlab{}.
\newblock \showarticletitle{The {James} construction and $\pi_4(\mathbb{S}^3)$
  in homotopy type theory}.
\newblock \bibinfo{journal}{\emph{Journal of Automated Reasoning}}
  \bibinfo{volume}{63} (\bibinfo{year}{2019}), \bibinfo{pages}{255--284}.
\newblock
\urldef\tempurl%
\url{https://doi.org/10.1007/s10817-018-9468-2}
\showDOI{\tempurl}
\showeprint{1710.10307}


\bibitem[Buchholtz and Hou(2020)]%
        {buchholtz2020cellular}
\bibfield{author}{\bibinfo{person}{Ulrik Buchholtz} {and}
  \bibinfo{person}{Kuen-Bang Hou}.} \bibinfo{year}{2020}\natexlab{}.
\newblock \showarticletitle{Cellular Cohomology in Homotopy Type Theory}.
\newblock \bibinfo{journal}{\emph{Logical Methods in Computer Science}}
  \bibinfo{volume}{16} (\bibinfo{year}{2020}).
\newblock
\urldef\tempurl%
\url{https://doi.org/10.23638/LMCS-16(2:7)2020}
\showDOI{\tempurl}
\showeprint{1802.02191}


\bibitem[Buchholtz and Rijke(2017)]%
        {buchholtz2017real}
\bibfield{author}{\bibinfo{person}{Ulrik Buchholtz} {and}
  \bibinfo{person}{Egbert Rijke}.} \bibinfo{year}{2017}\natexlab{}.
\newblock \showarticletitle{The real projective spaces in homotopy type
  theory}. In \bibinfo{booktitle}{\emph{Proceedings of the 32nd Annual ACM/IEEE
  Symposium on Logic in Computer Science (LICS)}}. IEEE, \bibinfo{pages}{1--8}.
\newblock
\urldef\tempurl%
\url{https://doi.org/10.1109/LICS.2017.8005146}
\showDOI{\tempurl}
\showeprint{1704.05770}


\bibitem[Champin et~al\mbox{.}(2024)]%
        {delooping-generated}
\bibfield{author}{\bibinfo{person}{Camil Champin}, \bibinfo{person}{Samuel
  Mimram}, {and} \bibinfo{person}{Émile Oleon}.}
  \bibinfo{year}{2024}\natexlab{}.
\newblock \bibinfo{title}{Delooping generated groups in homotopy type theory}.
  (\bibinfo{year}{2024}).
\newblock
\newblock
\shownote{Accepted at FSCD 2024}.


\bibitem[Cohen(2012)]%
        {cohen2012course}
\bibfield{author}{\bibinfo{person}{Marshall~M Cohen}.}
  \bibinfo{year}{2012}\natexlab{}.
\newblock \bibinfo{booktitle}{\emph{A course in simple-homotopy theory}}.
  Vol.~\bibinfo{volume}{10}.
\newblock \bibinfo{publisher}{Springer}.
\newblock
\urldef\tempurl%
\url{https://doi.org/10.1007/978-1-4684-9372-6}
\showDOI{\tempurl}


\bibitem[Hatcher(2009)]%
        {hatcher2009algebraic}
\bibfield{author}{\bibinfo{person}{Allen Hatcher}.}
  \bibinfo{year}{2009}\natexlab{}.
\newblock \bibinfo{booktitle}{\emph{Algebraic topology}}.
\newblock \bibinfo{publisher}{Cambridge University Press}.
\newblock


\bibitem[Hofmann and Streicher(1998)]%
        {hofmann1998groupoid}
\bibfield{author}{\bibinfo{person}{Martin Hofmann} {and}
  \bibinfo{person}{Thomas Streicher}.} \bibinfo{year}{1998}\natexlab{}.
\newblock \showarticletitle{The groupoid interpretation of type theory}.
\newblock \bibinfo{journal}{\emph{Twenty-five years of constructive type theory
  (Venice, 1995)}}  \bibinfo{volume}{36} (\bibinfo{year}{1998}),
  \bibinfo{pages}{83--111}.
\newblock
\urldef\tempurl%
\url{https://doi.org/10.1093/oso/9780198501275.003.0008}
\showDOI{\tempurl}


\bibitem[Kapulkin and Lumsdaine(2021)]%
        {kapulkin2021simplicial}
\bibfield{author}{\bibinfo{person}{Krzysztof Kapulkin} {and}
  \bibinfo{person}{Peter~LeFanu Lumsdaine}.} \bibinfo{year}{2021}\natexlab{}.
\newblock \showarticletitle{The simplicial model of Univalent Foundations
  (after Voevodsky)}.
\newblock \bibinfo{journal}{\emph{Journal of the European Mathematical
  Society}} \bibinfo{volume}{23}, \bibinfo{number}{6} (\bibinfo{year}{2021}),
  \bibinfo{pages}{2071--2126}.
\newblock
\urldef\tempurl%
\url{https://doi.org/10.4171/JEMS/1050}
\showDOI{\tempurl}
\showeprint{1211.2851}


\bibitem[Kraus and von Raumer(2022)]%
        {kraus2022rewriting}
\bibfield{author}{\bibinfo{person}{Nicolai Kraus} {and} \bibinfo{person}{Jakob
  von Raumer}.} \bibinfo{year}{2022}\natexlab{}.
\newblock \showarticletitle{A rewriting coherence theorem with applications in
  homotopy type theory}.
\newblock \bibinfo{journal}{\emph{Mathematical Structures in Computer Science}}
  \bibinfo{volume}{32}, \bibinfo{number}{7} (\bibinfo{year}{2022}),
  \bibinfo{pages}{982--1014}.
\newblock
\urldef\tempurl%
\url{https://doi.org/10.1017/S0960129523000026}
\showDOI{\tempurl}
\showeprint{2107.01594}


\bibitem[Licata and Finster(2014)]%
        {licata2014eilenberg}
\bibfield{author}{\bibinfo{person}{Daniel~R Licata} {and} \bibinfo{person}{Eric
  Finster}.} \bibinfo{year}{2014}\natexlab{}.
\newblock \showarticletitle{{Eilenberg-MacLane spaces in homotopy type
  theory}}. In \bibinfo{booktitle}{\emph{Proceedings of the 29th Annual
  ACM/IEEE Symposium on Logic in Computer Science (LICS)}}.
  \bibinfo{pages}{1--9}.
\newblock
\urldef\tempurl%
\url{https://doi.org/10.1145/2603088.2603153}
\showDOI{\tempurl}


\bibitem[Mac~Lane and Birkhoff(1999)]%
        {mac1999algebra}
\bibfield{author}{\bibinfo{person}{Saunders Mac~Lane} {and}
  \bibinfo{person}{Garrett Birkhoff}.} \bibinfo{year}{1999}\natexlab{}.
\newblock \bibinfo{booktitle}{\emph{Algebra}}. Vol.~\bibinfo{volume}{330}.
\newblock \bibinfo{publisher}{American Mathematical Society}.
\newblock


\bibitem[MacLane(2012)]%
        {maclane2012homology}
\bibfield{author}{\bibinfo{person}{Saunders MacLane}.}
  \bibinfo{year}{2012}\natexlab{}.
\newblock \bibinfo{booktitle}{\emph{Homology}}.
\newblock \bibinfo{publisher}{Springer Science \& Business Media}.
\newblock


\bibitem[Martin-L{\"o}f(1984)]%
        {martin1984intuitionistic}
\bibfield{author}{\bibinfo{person}{Per Martin-L{\"o}f}.}
  \bibinfo{year}{1984}\natexlab{}.
\newblock \bibinfo{booktitle}{\emph{Intuitionistic type theory}}.
  Vol.~\bibinfo{volume}{9}.
\newblock \bibinfo{publisher}{Bibliopolis Naples}.
\newblock


\bibitem[May(1999)]%
        {may1999concise}
\bibfield{author}{\bibinfo{person}{J~Peter May}.}
  \bibinfo{year}{1999}\natexlab{}.
\newblock \bibinfo{booktitle}{\emph{A concise course in algebraic topology}}.
\newblock \bibinfo{publisher}{University of Chicago press}.
\newblock


\bibitem[Milnor(1956a)]%
        {milnor1956construction}
\bibfield{author}{\bibinfo{person}{John Milnor}.}
  \bibinfo{year}{1956}\natexlab{a}.
\newblock \showarticletitle{Construction of universal bundles, {I}}.
\newblock \bibinfo{journal}{\emph{Annals of Mathematics}}
  (\bibinfo{year}{1956}), \bibinfo{pages}{272--284}.
\newblock


\bibitem[Milnor(1956b)]%
        {milnor1956construction2}
\bibfield{author}{\bibinfo{person}{John Milnor}.}
  \bibinfo{year}{1956}\natexlab{b}.
\newblock \showarticletitle{Construction of universal bundles, {II}}.
\newblock \bibinfo{journal}{\emph{Annals of Mathematics}}
  (\bibinfo{year}{1956}), \bibinfo{pages}{430--436}.
\newblock


\bibitem[Poincar{\'e}(1895)]%
        {poincare1895analysis}
\bibfield{author}{\bibinfo{person}{Henri Poincar{\'e}}.}
  \bibinfo{year}{1895}\natexlab{}.
\newblock \bibinfo{booktitle}{\emph{Analysis situs}}.
\newblock \bibinfo{publisher}{Gauthier-Villars Paris, France}.
\newblock


\bibitem[Rijke(2017)]%
        {rijke2017join}
\bibfield{author}{\bibinfo{person}{Egbert Rijke}.}
  \bibinfo{year}{2017}\natexlab{}.
\newblock \bibinfo{title}{The join construction}.  (\bibinfo{year}{2017}).
\newblock
\showeprint{1701.07538}
\newblock
\shownote{Preprint}.


\bibitem[Rijke(2018)]%
        {rijke2018classifying}
\bibfield{author}{\bibinfo{person}{Egbert Rijke}.}
  \bibinfo{year}{2018}\natexlab{}.
\newblock \emph{\bibinfo{title}{Classifying Types}}.
\newblock \bibinfo{thesistype}{Ph.\,D. Dissertation}. \bibinfo{school}{Carnegie
  Mellon University}.
\newblock
\showeprint{1906.09435}


\bibitem[Streicher(2018)]%
        {streicher2018fibered}
\bibfield{author}{\bibinfo{person}{Thomas Streicher}.}
  \bibinfo{year}{2018}\natexlab{}.
\newblock \bibinfo{title}{Fibered categories {\`a} la Jean Benabou}.
  (\bibinfo{year}{2018}).
\newblock
\showeprint{1801.02927}
\newblock
\shownote{Preprint}.


\bibitem[Tietze(1908)]%
        {tietze1908topologischen}
\bibfield{author}{\bibinfo{person}{Heinrich Tietze}.}
  \bibinfo{year}{1908}\natexlab{}.
\newblock \showarticletitle{{\"U}ber die topologischen Invarianten
  mehrdimensionaler Mannigfaltigkeiten}.
\newblock \bibinfo{journal}{\emph{Monatshefte f{\"u}r Mathematik und Physik}}
  \bibinfo{volume}{19} (\bibinfo{year}{1908}), \bibinfo{pages}{1--118}.
\newblock


\bibitem[{Univalent Foundations Program}(2013)]%
        {hottbook}
\bibfield{author}{\bibinfo{person}{The {Univalent Foundations Program}}.}
  \bibinfo{year}{2013}\natexlab{}.
\newblock \bibinfo{booktitle}{\emph{Homotopy Type Theory: Univalent Foundations
  of Mathematics}}.
\newblock \bibinfo{publisher}{\url{https://homotopytypetheory.org/book}},
  \bibinfo{address}{Institute for Advanced Study}.
\newblock
\showeprint{1308.0729}


\bibitem[W{\"a}rn(2023)]%
        {warn2023eilenberg}
\bibfield{author}{\bibinfo{person}{David W{\"a}rn}.}
  \bibinfo{year}{2023}\natexlab{}.
\newblock \bibinfo{title}{{Eilenberg-MacLane spaces and stabilisation in
  homotopy type theory}}.  (\bibinfo{year}{2023}).
\newblock
\showeprint{2301.03685}
\newblock
\shownote{Preprint}.


\bibitem[Watkins(1990)]%
        {watkins}
\bibfield{author}{\bibinfo{person}{Matthew Watkins}.}
  \bibinfo{year}{1990}\natexlab{}.
\newblock \bibinfo{title}{A Short Survey of Lens Spaces}.
  (\bibinfo{year}{1990}).
\newblock
\newblock
\shownote{Undergraduate dissertation}.


\end{thebibliography}

\onappendix{
\appendix

\section{Additional properties on deloopings}
We begin by detailing the proof of \cref{delooping-morphisms}.

\begin{proof}[Proof of \cref{delooping-morphisms}]
  We write $\Loop^{-1}f$ for the type of deloopings of~$f$, \ie the type of
  pointed maps $g:A\pto B$ whose looping is $f$, \ie
  \[
    \Sigma(g:A\to B).\Sigma(\pt g:g(\pt A)=\pt B).\Loop g=f
  \]
  %
  This type can be shown to be equivalent to the type
  \begin{equation}
    \label{eq:delooping-fun}
    \Sigma(g:A\to B).\Pi(a:A).C(a,g(a))
  \end{equation}
  where, for $a:A$ and $b:B$, the type $C(a,b)$ is defined as
  \[
    \Sigma(\pi:(a=\pt A)\to(b=\pt B)).D(a,\pi)
  \]
  with $D(a,\pi)$ defined as
  \begin{equation}
    \label{eq:delooping-D}
    \Pi(p:a=\pt A).\Pi(q:\pt A=\pt A).\pi(p\pcomp q)=\pi(p)\pcomp f(q)
  \end{equation}
  see~\cite[Lemma~9]{warn2023eilenberg}.
  This can be explained as follows. Any delooping~$g:A\pto B$ of~$f$ induces a
  function $\pi:(a=\pt A)\to(b=\pt B)$, defined by $\pi(p)=\ap g p\pcomp\pt g$,
  which satisfies, for $p:a=\pt A$ and $q:\pt A=\pt A$,
  \begin{align*}
    \pi(p\pcomp q)
    &=\ap{g}{p\pcomp q}\pcomp\pt g\\
    &=\ap{g}p\pcomp\ap{g}q\pcomp\pt g\\
    &=\ap{g}p\pcomp\pt g\pcomp\Loop g(q)\\
    &=\pi(p)\pcomp\Loop g(q)
  \end{align*}
  and thus the type~$C(a,g(a))$ is inhabited for any~$a:A$. The above
  equivalence shows that this characterizes the deloopings of~$f$: those are
  precisely the functions~$g$ such that, for any $a:A$, there exists a
  function~$\pi$ satisfying the above property.

  From the equivalence with \cref{eq:delooping-fun}, by type-theoretic choice,
  $\Loop^{-1}f$ is also equivalent to
  \begin{equation}
    \label{eq:delooping-sigma}
    \Pi(a:A).\Sigma(b:B).C(a,b)
  \end{equation}
  Because $B$ is a groupoid, the last equality in the definition
  \cref{eq:delooping-D} of $D(a,\pi)$ is a proposition and thus also $D(a,\pi)$
  itself. We show below that we have, for any $b:B$,
  \begin{equation}
    \label{eq:delooping-Cpt}
    C(\pt A,b)\qeq(b=\pt A)
  \end{equation}
  From there we deduce that $\Sigma(b:B).C(\pt A,b)$ is equivalent to
  $\Sigma(b:B).(b=\pt A)$ and is thus contractible. By path induction, we thus
  have
  \[
    \Pi(a:A).(a=\pt A)\to\isContr(\Sigma(b:B).C(a,b))
  \]
  and thus
  \[
    \Pi(a:A).\ptrunc{a=\pt A}\to\isContr(\Sigma(b:B).C(a,b))
  \]
  because being contractible is a proposition. Since $A$ is connected, we deduce
  that $\Sigma(b:B).C(a,b)$ is contractible for every $a:A$. The type
  \cref{eq:delooping-sigma} is thus contractible and thus also $\Loop^{-1}f$,
  which is what we wanted to show.

  We are left with showing \cref{eq:delooping-Cpt}. It can be shown that for any
  suitably typed function~$\pi$, the type $D(\pt A,\pi)$ is equivalent to
  \[
    \Pi(q:\pt A=\pt A).\pi(q)=\pi(\refl)\pcomp f(q)
  \]
  Namely the former implies the later as a particular case and, conversely,
  supposing the second one, we have for $p:\pt A=\pt A$ and $q:\pt A=\pt A$,
  \[
    \pi(p\pcomp q)
    =
    \pi(\refl)\pcomp f(p\pcomp q)
    =
    \pi(\refl)\pcomp f(p)\pcomp f(q)
    =
    \pi(p)\pcomp f(q)
  \]
  because~$f$ preserves composition. From there follows easily
  \cref{eq:delooping-Cpt}, \ie that
  \[
    \Sigma(\pi:(\pt A=\pt A)\to(b=\pt B)).\pi(q)=\pi(\refl)\pcomp f(q)
  \]
  is equivalent to $b=\pt B$, since, in the above type, the second component
  expresses that the function $\pi$ in the first component is uniquely
  determined by $\pi(\refl)$, which is an element of $b=\pt B$. A similar
  approach is developed in~\cite[Section~4.10]{symmetry}. This result also
  follows from~\cite[Corollary~12]{warn2023eilenberg}.
\end{proof}

\begin{lemma}
  \label{delooping-functorial}
  The delooping of morphisms is functorial: given two morphisms
  $f:\Loop A\to\Omega B$ and $g:\Omega B\to\Omega C$, we have
  $\B(g\circ f)=\B g\circ \B f$ and $\B\id{\Loop A}=\id{A}$.
\end{lemma}
\begin{proof}
  By functoriality of $\Loop$, we have
  $\Loop(\B g\circ \B f)=\Omega\B g\circ\Omega\B f$ and thus
  $\B(g\circ f)=\B g\circ \B f$ by \cref{delooping-morphisms}. Preservation of
  identities can be proved similarly.
\end{proof}

The following lemma states that deloopings of groups are unique, in a strong
sense. We write $\PCGpd$ for the type of pointed connected groupoids.

\begin{lemma}
  \label{delooping-unique}
  Given a group~$G$, type $\Sigma(A:\PCGpd).(\Loop A=G)$ is contractible. In
  particular, for any two deloopings $\B G$ and $\B' G$, we have $\B G=\B' G$.
\end{lemma}
\begin{proof}
  Suppose given two deloopings of~$G$, by which we mean two types $A,A':\PCGpd$ equipped
  with identities $p:\Loop A=G$ and $q:\Omega A'=G$. We thus have a morphism
  $f:\Loop A\to\Omega A'$, defined as $f:=\transportinv{q}\circ\transport{p}$,
  which induces, by \cref{delooping-morphisms}, a morphism $\B f:A\to A'$ such
  that $\Loop\B f=f$. By \cref{delooping-morphisms}, $\B f$ is an isomorphism
  and thus induces an identity $A=A'$ by univalence.
  We finally show that $\Sigma(A:\PCGpd).(\Loop A=G)$ is a proposition. Given
  two elements $(A,p)$ and $(B,q)$ of this type, by \cite[Theorem 2.7.2 and
  Theorem 2.11.3]{hottbook}, an equality between them consists of an equality
  $r:A=B$ such that $\sym{\Loop r}\pcomp p=q$. This amounts to show
  $\transport p=\transport{\Loop r}\circ\transport q$, \ie
  $\transport p=\Loop\B f\circ\transport q$, which follows easily from the
  equality $\Loop\B f=f$.
\end{proof}

\section{Delooping with torsors}
\label{delooping-with-G-sets}
We recall here a classical torsor construction in order to define the delooping
of a group~$G$.
A \emph{$G$-type} is a type~$A$ together with a (left) action of~$G$ on~$A$, \ie
a morphism $G\to\Aut(A)$ to the group of automorphisms of~$A$. We write
$g\cdot a$ for the action of an element of $g:G$ on $a:A$. A \emph{$G$-set} is
the particular case of an action of~$G$ on a set $A$. Any group~$G$ acts on
itself by multiplication, and thus canonically induces a $G$-set $P_G$ called
the \emph{principal $G$-set}. A morphism of $G$-sets is a morphism $f:A\to B$ between
the underlying sets which preserves the action:
$f(g\cdot a)=g\cdot f(a)$ for every $g:G$ and $a:A$. We write $\GSet$ for the
type of $G$-sets, $A\to_GB$ for the type of morphisms of $G$-sets and
$A\simeq_G B$ for the type of isomorphisms of~$G$-sets.

The main property of connected components is the following one.

\begin{lemma}
  \label{connected-component-embedding}
  For any type~$A$ with a distinguished element~$\pt{}$, the type
  $\Comp(A,\pt{})$ is connected and the first projection
  $\iota:\Comp(A,\pt{})\to A$ is an embedding.
\end{lemma}
\begin{proof}
  Easy, left to the reader.
\end{proof}

We can then show any group admits a delooping (\cref{deloopings-exist}) as
follows:

\begin{theorem}
  Every group~$G$ admits a delooping~$\B G$, which can be defined as
  $\B G=\Comp(\GSet,P_G)$.
\end{theorem}
\begin{proof}
  By \cref{connected-component-embedding}, it is enough to show that
  $(P_G=P_G)=G$. By univalence, see~\cite[Section~2.14]{hottbook}, the type
  $P_G=P_G$ is equivalent to $P_G\simeq_G P_G$, and we are left with showing
  that it is in turn equivalent to~$G$. Any element $g:G$ induces an isomorphism
  $[g]:P_G\to_G P_G$ defined by $[g](h)=hg$. Conversely, to any isomorphism
  $f:P_G\to_G P_G$, we associate the element $f(1):G$. The two operations are
  mutually inverse group morphisms: given $g:G$, we have $[g](1)=1g=g$; given an
  isomorphism $f:P_G\to_G P_G$, we have $f(g)=f(g1)=g f(1)=[f(1)](g)$. Thus
  inducing the required equivalence.
\end{proof}

The fundamental theorem of deloopings
(\cref{groups-vs-pointed-connected-groupoids}) can then be shown as follow for
this model:

\begin{theorem}
  The functions $\Loop$ and $\B$ induce an equivalence between the type of
  groups and the type of pointed connected groupoids.
\end{theorem}
\begin{proof}
  We have seen that, for any type~$A$, $\Loop A$ has a canonical group
  structured induced by path composition and constant paths. The type of sets is
  a groupoid~\cite[Theorem 7.1.11]{hottbook}, and given a group~$G$ and a
  set~$A$, the type of morphisms $G\to\Aut(A)$ is a set and thus a groupoid;
  therefore the type of $G$-sets is a groupoid~\cite[Theorem 7.1.8]{hottbook},
  and thus also $\B G$ by \cref{connected-component-embedding}. Given a
  group~$G$, we have $\Loop\B G=G$ by definition of deloopings. Given a pointed
  connected groupoid~$A$, $A$ and $\B\Loop A$ are both deloopings of~$\Omega A$
  (with the canonical identities $\Loop A=\Omega A$ and
  $\Loop\B\Omega A=\Omega A$) and are thus equal by \cref{delooping-morphisms}.
\end{proof}

Finally, we show that external and internal group actions on sets coincide:

\begin{lemma}
  \label{internal-group-action}
  We have $\GSet=(\B G\to\HSet)$.
\end{lemma}
\begin{proof}
  Fix a set~$A$. The type $\HSet$ is a groupoid~\cite[Theorem 7.1.11]{hottbook}
  and therefore, by \cref{connected-component-embedding}, $\Comp(\HSet,A)$ is a
  pointed connected groupoid. By \cref{groups-vs-pointed-connected-groupoids},
  actions on~$A$, \ie morphisms of groups
  $\Loop\B G\to\Omega\Comp(\HSet,A)$
  correspond to morphisms of pointed connected
  groupoids $\B G\pto\Comp(\HSet,A)$. The result follows by summing the
  correspondence over~$A:\HSet$.
\end{proof}

\noindent
Following what we have explained in \cref{delooping-with-G-sets}, the above
correspondence can be interpreted as the fact that \emph{external} actions
(elements of $\GSet$) correspond to \emph{internal} ones (morphisms in
$\B G\to\HSet$).

\section{Proof of \cref{non-prime-fiber-sequence}}
\label{non-prime-fiber-sequence-proof}
The proof below requires the following classical result, called the
\emph{flattening lemma}~\cite[Section~6.12]{hottbook}, which we first recall (a
variant of this lemma for pushouts was given in \cref{flattening-pushout}):

\begin{lemma}[Flattening for coequalizers]
  \label{flattening-coequalizers}
  Suppose given a coequalizer
  \[
    \begin{tikzcd}
      A\ar[r,shift left,"f"]\ar[r,shift right,"g"']&B\ar[r,dotted,"h"]&C
    \end{tikzcd}
  \]
  with $p:h\circ f=h\circ g$, and a type family $P:C\to\U$. Then the diagram
  \[
    \begin{tikzcd}[sep=huge]
      \Sigma A.(P\circ h\circ f)\ar[r,shift left,"\Sigma f.(\lambda\_.\id{})"]\ar[r,shift right,"\Sigma g.e"']&\Sigma B.(P\circ h)\ar[r,dotted,"\Sigma h.(\lambda\_.\id{})"]&\Sigma C.P
    \end{tikzcd}
  \]
  is a coequalizer, where the map
  \[
    e:(a:A)\to P(h(f(a)))\to P(h(g(a)))
  \]
  is induced by transport along~$p$ by
  \[
    e\,a\,x\defd
    \subst{P}{\happly pa}(x)
    \text.
  \]
\end{lemma}

\noindent
Note that there is a slight asymmetry: we could have formulated a similar
statement with $\Sigma A.(P\circ h\circ g)$ as left object.

We can now prove \cref{non-prime-fiber-sequence} as follows.

\begin{proof}[Proof of \cref{non-prime-fiber-sequence}]
  Let us compute the fiber of the map~$\B\phi^l$, which is
  \[
    \B\phi^l\defd\Sigma(x:\S1).(\pt{}=\B\phi^l(x))
  \]
  We write $P:\S1\to\U$ for the map defined by
  \[
    P(x)\defd(\pt{}=\B\phi^l(x))
  \]
  whose total space is the above fiber. The sphere $\S1$ can be obtained as the
  coequalizer
  \[
    \begin{tikzcd}
      1\ar[r,shift left,"f"]\ar[r,shift right,"g"']&1\ar[r,dotted,"h"]&\S1
    \end{tikzcd}
  \]
  where $h(\pt{})=\pt{}$ is the base point of the circle and the equality
  $p:h\circ f=h\circ g$ is such that $\happly p{\pt{}}:\pt{}=\pt{}$ is the
  canonical non-trivial loop of the circle, which is denoted~$q$ below. The
  flattening lemma for coequalizers (\cref{flattening-coequalizers}) ensures
  that that total space of~$P$ is the following coequalizer
  \[
    \begin{tikzcd}[sep=huge]
      \Sigma 1.(P\circ h\circ f)\ar[r,shift left,"\Sigma f.(\lambda\_.\id{})"]\ar[r,shift right,"\Sigma g.e"']&\Sigma 1.(P\circ h)\ar[r,dotted,"\Sigma h.(\lambda\_.\id{})"]&\Sigma\S1.P
    \end{tikzcd}
  \]
  where the map $e$ is induced by transport along~$p$, \ie
  \[
    e\,x\,r
    \defd
    \subst P{\happly px}(r)
    \text.
  \]
  By using the fact that for every $A:1\to\U$, we have
  \[
    \Sigma(x:1).A(x)=A(\pt{})
  \]
  the above coequalizer can be rewritten as
  \[
    \begin{tikzcd}
      P(\pt{})\ar[r,shift left,"\id{}"]\ar[r,shift right,"e(\pt{})"']&P(\pt{})\ar[r,dotted]&\Sigma\S1.P
    \end{tikzcd}
  \]
  with, for $r:\pt{}=\B\phi^l(\pt{})$ in $\B\Z_m$,
  \begin{align*}
    e(\pt{})(r)
    &=\subst P{\happly p{\pt{}}}(r)&&\text{by definition of~$e$}\\
    &=\subst Pq(r)&&\text{by definition of~$p$}\\
    &=r\pcomp\ap{(\B\phi^l)}q&&\text{by path transport~\cite[Theorem~2.11.4]{hottbook}}    
  \end{align*}
  Moreover, we have
  \begin{align*}
    P(\pt{})&=(\pt{}=\B\phi^l(\pt{}))&&\text{by definition of~$P$}\\
    &=(\pt{}=\pt{})&&\text{because $\B\phi^l$ is pointed}\\
    &=\Loop\B\Z_m&&\text{by definition of $\Loop$}\\
    &=\Z_m&&\text{because $\B\Z_m$ is a delooping of~$\Z_m$}
  \end{align*}
  and the above equalizer can be rewritten as
  \[
    \begin{tikzcd}[sep=large]
      \Z_m\ar[r,shift left,"\id{}"]\ar[r,shift right,"\lambda i.(i+l)"']&\Z_m\ar[r,dotted]&\Sigma\S1.P
    \end{tikzcd}
  \]
  Finally, it can be decomposed as the coproduct of~$k$ copies of the coequalizer
  \[
    \begin{tikzcd}[sep=large]
      \Z_{m/k}\ar[r,shift left,"\id{}"]\ar[r,shift right,"\lambda i.(i+1)"']&\Z_{m/k}\ar[r,dotted]&\S1
    \end{tikzcd}
  \]
  with $k=\pgcd(l,m)$. Namely, we have an isomorphism
  \begin{align*}
    f:\Z_m&\to k\Z_{m/k}\\
    i&\mapsto(i\mod k,i/k)
  \end{align*}
  (an element of $k\Z_{m/k}$ can be seen as a pair $(i,j)$ with $0\leq i<k$ and
  $j\in\Z_{m/k}$), whose inverse sends $(i,j)$ to $kj+i$, which makes the diagram
  \[
    \begin{tikzcd}[sep=large]
      \Z_m\ar[d,"f"']\ar[r,shift left,"\id{}"]\ar[r,shift right,"\lambda i.(i+l)"']&\Z_m\ar[d,"f"]\\
      k\Z_{m/k}\ar[r,shift left,"k\id{}"]\ar[r,shift right,"k(\lambda i.i+1)"']&k\Z_{m/k}
    \end{tikzcd}
  \]
  commute.
\end{proof}

}
\end{document}